\documentclass{aa}  
\usepackage{graphicx}
\usepackage{txfonts}
\usepackage{ulem}
\usepackage{graphicx}
\usepackage{txfonts}
\usepackage{longtable}
\usepackage{verbatim}
\usepackage{tablefootnote}
\usepackage{url}
\usepackage{float,color}
\usepackage{subcaption}
\usepackage[figuresright]{rotating}
\usepackage{natbib}
\usepackage{hyperref}
\hypersetup{
    colorlinks=true,
    linkcolor=blue,
    filecolor=magenta,      
    urlcolor=cyan,
    citecolor=blue
}
\newcommand{\vmic}{\xi_{\rm t}}
\newcommand{\vmac}{V_{\rm mac}}

\newcommand{\opd}{\log \tau_{\rm 5000\text{\AA}}}
\newcommand{\td}{\langle {\rm 3D} \rangle}
\newcommand{\teff}{T_{\rm eff}}
\newcommand{\feh}{\rm{[Fe/H]}}
\newcommand{\logg}{\log{g}}

\newcommand{\ha}{H$\alpha$}
\begin{document} 

   \title{Non-LTE Radiative Transfer with Turbospectrum}

   \subtitle{}

   \author{Jeffrey M. Gerber
          \inst{\ref{inst:MPIA}}
          \and
          Ekaterina Magg\inst{\ref{inst:MPIA}}
          \and
          Bertrand Plez\inst{\ref{inst:montp}}
          \and
          Maria Bergemann\inst{\ref{inst:MPIA}}
          \and
          Ulrike Heiter\inst{\ref{inst:Uppsala}}
          \and
          Terese Olander\inst{\ref{inst:Uppsala}}
          \and
          Richard Hoppe\inst{\ref{inst:MPIA}}
          }

   \institute{Max Planck Institute for Astronomy,
              69117, Heidelberg, Germany\\
              \email{gerber@mpia.de}
              \label{inst:MPIA}
         \and
             LUPM, Université de Montpellier, CNRS, Montpellier, France
             \label{inst:montp}
        \and
             Observational Astrophysics, Department of Physics and Astronomy, Uppsala University, Box 516, 751 20 Uppsala, Sweden
             \label{inst:Uppsala}
             }

   \date{Received xxxx; accepted xxxx}

 
  \abstract{Physically realistic models of stellar spectra are needed in a variety of astronomical studies, from the analysis of fundamental stellar parameters, to studies of exoplanets and stellar populations in galaxies. Here we present a new version of the widely-used radiative transfer code Turbospectrum, which we update with the capacity to perform spectrum synthesis for lines of multiple chemical elements in Non-Local Thermodynamic Equilibrium (NLTE). We use the code in the analysis of metallicites and abundances of the Gaia FGK benchmark stars, using one-dimensional MARCS atmospheric models and the averages of 3D radiation-hydrodynamics simulations of stellar surface convection. We show that the new more physically realistic models offer a better description of the observed data and make the program and the associated microphysics data publicly available, including grids of NLTE departure coefficients for H, O, Na, Mg, Si, Ca, Ti, Mn, Fe, Co, Ni, Sr, and Ba.}

   \keywords{Radiative transfer - Methods: observational - Techniques: spectroscopic - Sun: abundances - Stars: abundances}

   \maketitle
%

\section{Introduction}
Diagnostic stellar spectroscopy is one of the major instruments in modern astrophysical research. With the advent of large-scale stellar surveys, such as Gaia-ESO, APOGEE, SDSS, GALAH, Gaia, 4MOST, and WEAVE, we are entering a new era of precision stellar physics. Large sets of high quality stellar spectra can be exploited to derive direct observational constraints on stellar chemical composition, and therefore on the detailed chemical evolution of stellar populations and the Milky Way Galaxy.

In this study, we focus on the role of physical processes in stellar spectroscopy, in particular on the physics of non-local thermodynamic equilibrium (NLTE) in the determination of stellar metallicities and detailed abundances. Much of the previous work in the field has been carried out under the assumption of local thermodynamic equilibrium (LTE), in which the atomic number densities have been computed using the Saha-Boltzmann formulae from statistical mechanics. However, in the atmospheres of FGKM-type stars, the complex interaction of strong highly non-local radiation fields and the gas particles leads to deviations from LTE. This detailed microphysics can nowadays be modeled using the concept of statistical equilibrium \citep[e.g.][]{Asplund2005, Mashonkina2007, Lind2011, Bergemann2012, Amarsi2016, Semenova2020, Masseron2021}, in which chemical elements are modelled under the assumption of a trace element, that is including the NLTE effects in the line formation and spectrum synthesis, but ignoring their influence on the energy balance and thus, the thermodynamic structure of stellar atmospheres. This assumption has, so far, been used in all studies, in which NLTE modelling is used to determine abundances for large stellar samples \citep{Kovalev2019,Buder2020,Amarsi2020}. 

Another concept in classical stellar spectroscopy is the assumption of one-dimensional (1D) geometry and hydrostatic equilibrium (HE). Different families of 1D HE LTE atmospheric models have been computed, including Phoenix \citep{Hauschildt1999}, ATLAS \citep{Kurucz1979, Kurucz2005}, MARCS \citep{Gustafsson2008}, and MAFAGS-OS \citep{Grupp2004a,Grupp2004b}. However, despite their general utility and low computational cost, the models offer only an approximate description of stellar atmospheres, as convection and turbulence -- critical physical processes in stars with convective envelopes -- are replaced by highly-simplified parametrisations \citep[e.g][]{Nordlund2009,Freytag2012}. Convective energy transport is typically modelled using the mixing-length theory \citep{Bohm1958} and turbulence is represented using the ``macro-'' and ``microturbulence'' parameters. The classical 1D HE models have the advantage of including a very detailed account of opacities, although the opacities are still computed in LTE.

The main goal of this paper is to introduce an open-source NLTE version of the widely-used Turbospectrum synthesis code \citep[TS,][]{Alvarez1998,Plez2012}\footnote{New NLTE version is presented at: \url{https://github.com/bertrandplez/Turbospectrum2020}. The previous LTE release 19.1.4 is available at: \url{https://github.com/bertrandplez/Turbospectrum2019}.}. Our main motivation is to provide a public software that allows the user to generate realistic wide-band synthetic stellar spectra of FGKM-type stars relying on state-of-the-art NLTE model atoms for multiple chemical elements simultaneously and at low computational cost, comparable to standard 1D LTE codes, like MOOG \citep{Sneden1973} or Synthe \citep{Kurucz1970}. The code can self-consistently handle complex blends, hyperfine structure, isotopic shifts, and enables treatment of very large - multi-million - linelists, including all atoms and tens of molecules. We also provide associated ready-to-use 1D and average 3D model atmospheres in the appropriate format and the grids of NLTE departure coefficients. The code is conceptually similar to the SIU \citep{SIUReetz,Reetz1999}, SME \citep{sme,Piskunov2017}, and SYNSPEC \citep{Hubeny2021} codes, which operate, in the NLTE case, with pre-computed grids of NLTE departure coefficients\footnote{A departure coefficient describes the ratio of the number density of an atom at a given energy level computed in NLTE to that computed in LTE, i.e. using the Saha-Boltzmann equilibrium relations.}\footnote{SYNSPEC can also work with NLTE populations.}. However, the NLTE versions of the former codes are not open-source. Also, different from SIU, the NLTE TS code feeds on a more comprehensive and up-to-date set of background bound-free and bound-bound opacities. Furthermore, it is superior to SME in terms of computational efficiency, as a detailed high-resolution spectrum from 400 to 900 nm of a typical G-type solar-metallicity star can be computed on timescales of a few minutes. For example, we were able to generate a 1D NLTE spectrum of the Sun with this wavelength range and a sampling of 0.05 \AA~on a 2.5GHz processor computing roughly 35 million atomic and molecular lines in 12 minutes. This model also included NLTE computations for all 13 elements for which we have pre-computed departure coefficient grids. The most recent version of SYNSPEC also offers a capability to calculate NLTE synthetic spectra, but NLTE is included in lines of H, Ca, and Mg \citep{Hubeny2021} only.

The paper is organised as follows. Sect. \ref{sec:methodology} describes Turbospectrum (hereafter, TS) and the model atmospheres used. We then outline the updates to TS in Sect. \ref{sec:newturbo}, comment on the comparison with another radiative transfer code (Sect. \ref{sec:tests}), demonstrate the capability to compute NLTE calculations for multiple elements at once (Sect. \ref{sec:nltemultiple}) and the ability to compute 1D and $\td$ NLTE H line profiles (Sect. \ref{sec:hprofiles}). In Sect. \ref{sec:exampleAnalysis}, we provide an example of the type of abundance analysis that can be done with TS using observations of Gaia benchmark stars as a sample. Finally, we present the results of our example analysis for Fe and Ca in Sect. \ref{sec:abundResults}, and we discuss our findings in the context of previous studies in the literature in Sect. \ref{sec:discussion}.
\section{Methodology}\label{sec:methodology}
\subsection{Model atmospheres}\label{sec:atmos}

We make use of two grids of stellar model atmospheres computed for low-mass late-type (FGKM) stars. We use the 1D line-blanketed hydrostatic MARCS models \citep{Gustafsson2008} and average 3D Stagger models (hereafter, $\td$) from \citet{Magic2013a,Magic2013b}. We briefly describe the models below. 

\begin{figure}[h!]
\centering
\includegraphics[width=\linewidth]{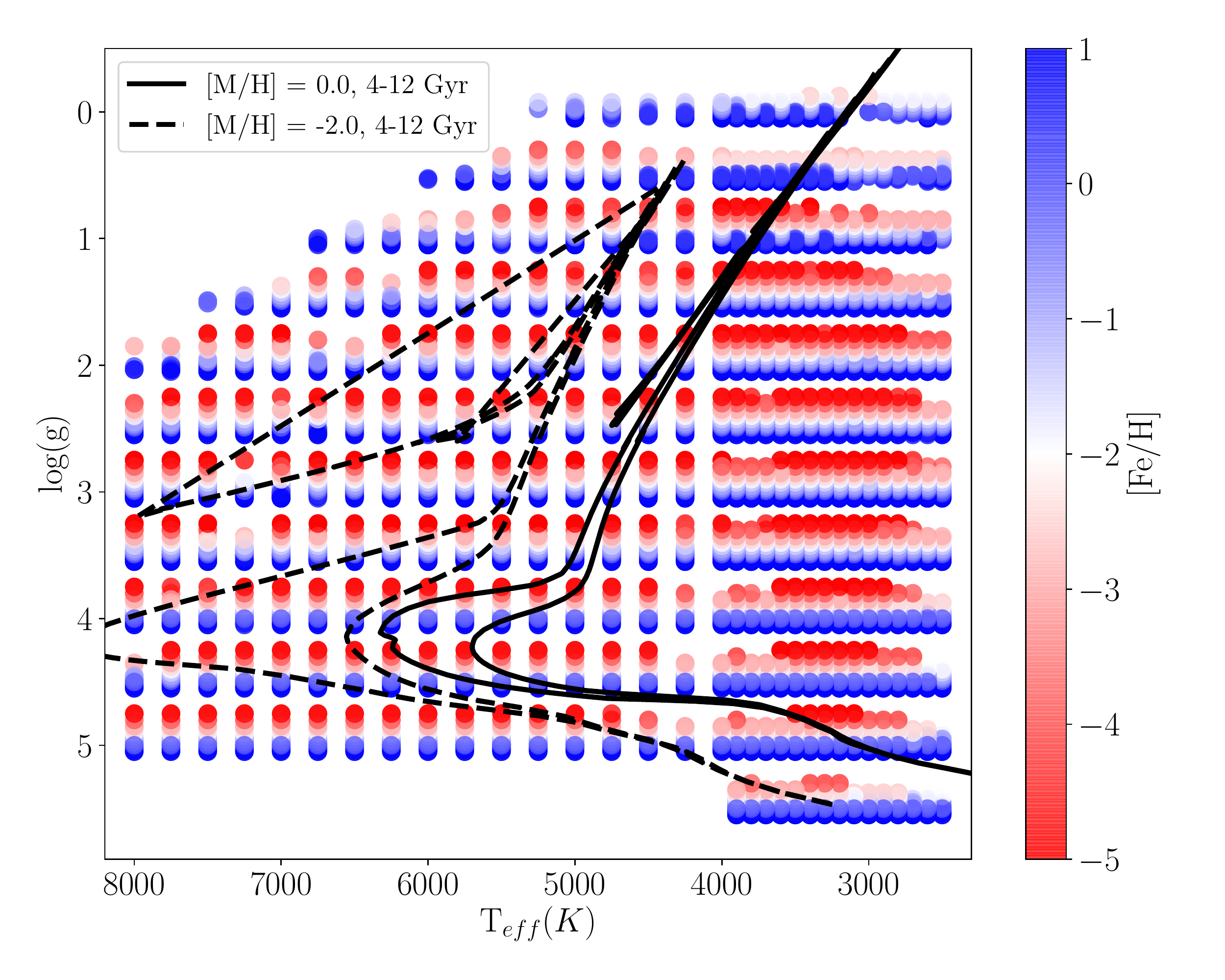}
\caption{MARCS model atmosphere grid used in this work. PARSEC isochrones \citep{Bressan2012} calculated for [M/H] = -2 and 0 are overplotted as dashed and solid black lines, respectively.}
\label{fig:MARCS_NLTE_grids_Space}%
\end{figure}

The MARCS\footnote{\url{https://marcs.astro.uu.se}} models were computed in LTE, solving the equation of radiative transfer in plane-parallel geometry for dwarf models (surface gravity $\log g [{\rm cm~s}^{-2}] \ge 3.0$) and in spherical symmetry for giants ($\log g \le 3.5$). Convective energy transport is computed in the framework of the mixing-length theory \citep{Henyey1965}, with the mixing-length coefficient $\alpha$ set to $1.5$. Very comprehensive treatment of radiative bound-bound and bound-free opacities, including all atoms in the first and second ionisation stages, as well as $519$ di- and tri-atomic molecules, are included in the code. The grid relies on the solar abundance mixture by \citet{Grevesse2007}, however, the abundances of oxygen and $\alpha$-capture-elements are enhanced for models with metallicity [Fe/H] $\le-0.25$, reflecting the typical elemental abundance ratios in stars as a function of metallicity in the solar neighbourhood\footnote{The enhancement is +0.1 for [Fe/H]=$-0.25$, +0.2 for [Fe/H]=$-0.5$, +0.3 for [Fe/H]=$-0.75$, and +0.4 for [Fe/H]$\le-1.0$.}. The parameter space of the MARCS grid is shown in Fig.~\ref{fig:MARCS_NLTE_grids_Space}. The grids cover the following range of stellar parameters: $\teff = 2500, 8000$ K with a step of 100 K from 2500 to 4000 K and 250 K from 4000 to 8000 K, $\log g = 0, 5.5$ dex with a step of 0.5 dex, [Fe/H] $= -5.0, 1.0$ with a step of 1 dex from -5 to -3, 0.5 dex from -3 to -1, and 0.25 dex from -1 to 1 dex. The grids also include three different values for microturbulence at 1, 2, and 5 km/s. 

The $\td$ Stagger models were adopted from \citet{Magic2013a,Magic2013b}\footnote{ \url{https://staggergrid.wordpress.com}}. These models were constructed by spatial and temporal averaging of 3D radiation-hydrodynamics (RHD) simulations of stellar surface convection. Specifically, averages were carried out on surfaces of equal optical depth (log$\tau_{5000}$). The Stagger code relies on the modified version of the equation of state (EoS) from \citet{Mihalas1988} and abundances were taken from \citet{Asplund2009}. For a more detailed description of the input physics and technical aspects, we refer the reader to \citet{Magic2013a,Magic2013b}. Since in these simulations, convection and turbulence are the natural consequence of solving the equations of fluid dynamics coupled with the equation of radiative transfer, there is no need for introducing ad-hoc free parameters, such as the mixing length or  ``microturbulence'' $\vmic$, which directly affects the line opacity. Therefore, we include a depth-dependent velocity profile computed from the original 3D velocity field in the simulation cubes in the form of a microturbulence with a depth-dependent value of one standard deviation of the 3D components as suggested in \citet{Uitenbroek2011}.

It has not been extensively tested yet, whether this approximation provides the most realistic description of turbulent flows in stellar atmospheres. For some elements, like Mg \citep{Bergemann2017} the $\td$ NLTE approach works well, whereas for the other species like Fe \citep{Amarsi2016} the differences between $\td$ NLTE and full 3D NLTE are somewhat larger. Nonetheless, the $\td$ model atmosphere approach has an important advantage over standard 1D hydrostatic model atmospheres, in that it eliminates two important free parameters (mixing length and $\vmic$) from the modelling. Therefore and most crucially, our $\td$ spectral models are predictive. Especially the classical $\vmic$ is, numerically, defined as a simple correction to the line opacity, which is commonly either set by hand, or is optimised using some recipes defined by each user. Thus, it is partially degenerate with the abundances of chemical elements. This implies, strictly speaking, that the user has the freedom to influence the resulting chemical abundance estimate obtained by fitting standard 1D HE synthetic spectra to observed data.

\begin{figure*}[h!]
\centering
\includegraphics[scale=0.45]{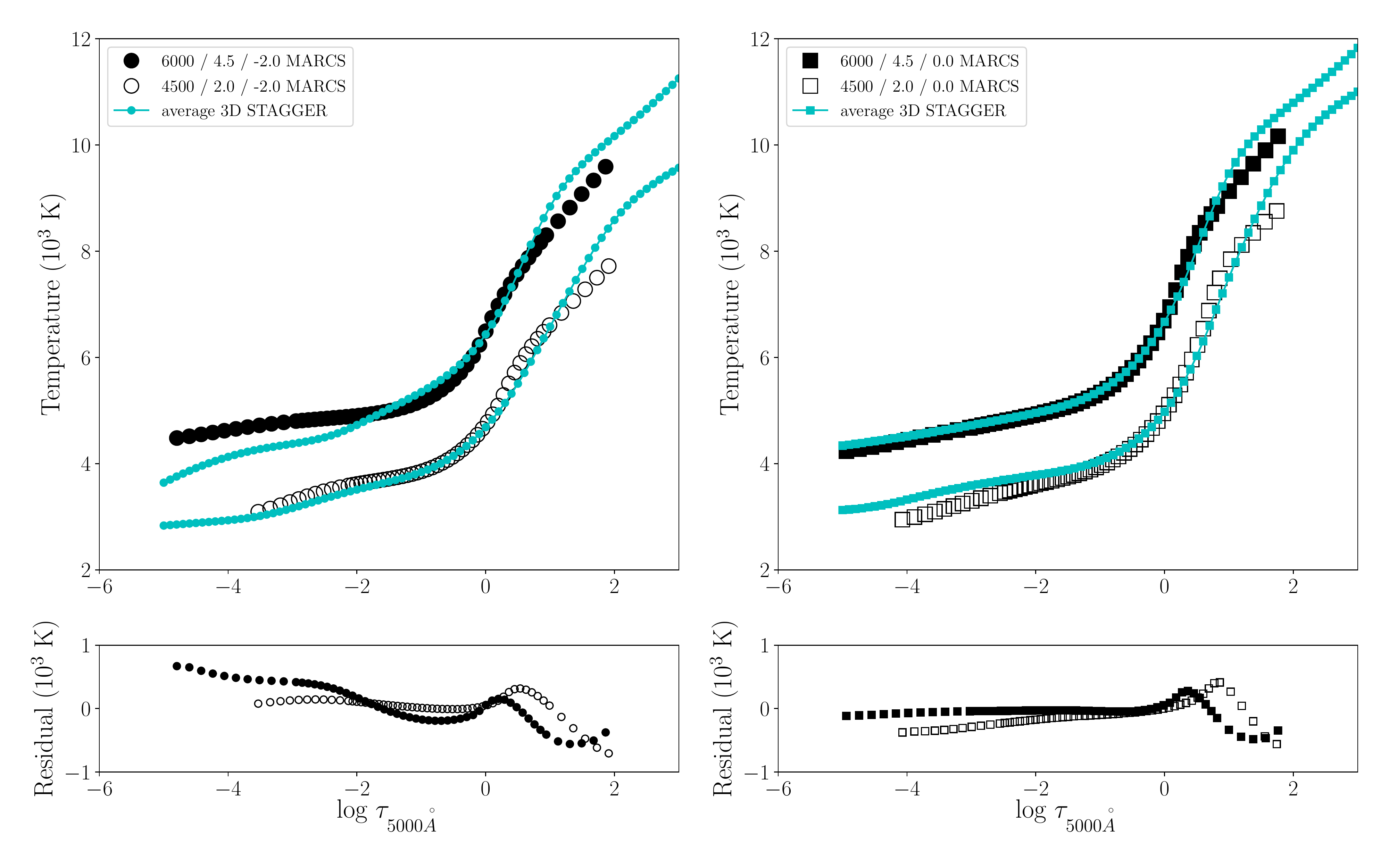}
\caption{The top two panels show temperature as a function of optical depth at 5000 \AA\ for model atmospheres of four types of stars. The open (giant stars) and solid (dwarf stars) black circles and squares show the MARCS model atmospheres for the parameters given in the legend (low metallicity to the left, solar metallicity to the right). The corresponding $\td$ Stagger models are shown as cyan dots. The bottom two panels show the difference between the MARCS model and the $\td$ Stagger model for each case following the same convention for open and solid circles or squares as in the top panels.}
\label{fig:TeffTauMARCSvsSTAGGER}
\end{figure*}

Figure~\ref{fig:TeffTauMARCSvsSTAGGER} compares the structures of the MARCS and $\td$ Stagger models by showing temperature as a function of optical depth at 5000~\AA\ ($\opd$) for several model atmospheres representative of low mass dwarfs (effective temperature $\teff = 6000$~K, $\logg = 4.5$) and giants ($\teff = 4500$~K, $\logg = 2.0$) at low and solar metallicity ($\feh = -2$ and 0, respectively). We also show the difference between the MARCS model and the corresponding $\td$ Stagger model. As can be seen, the differences between the 1D and $\td$ models are in general largest in the interior. In the case of the low metallicity dwarf the outer layers of the atmosphere also show significant differences almost reaching 1000~K.
\subsection{Turbospectrum}\label{sec:turbos}
Turbospectrum \citep{Alvarez1998,Plez2012} is an LTE radiative transfer code based on the same methods and input physics as the MARCS model atmosphere code. It uses the Feautrier scheme \citep{Feautrier1964, Nordlund1984} to solve the radiative transfer equation including scattering, and works in plane-parallel and spherically symmetric geometry for flux and intensities at various angles. Rayleigh and electron scattering in the continuum is fully taken into account, although we note that this effect is mostly relevant in the atmospheres of extremely metal-poor red giants and it influences only the spectra in the near-UV \citep{Gustafsson1975,Cayrel2004,Sobeck2011}. TS uses exactly the same EoS as MARCS with up to four ions of all $92$ natural elements, and hundreds of molecules and radicals. Continuous opacities are identical in both codes, and they include bound-free transitions for all major species (\ion{H}{i}, H$^-$, \ion{He}{i}, \ion{C}{i}, \ion{C}{ii}, \ion{N}{i}, \ion{N}{ii}, \ion{O}{i}, \ion{O}{ii}, \ion{Mg}{i}, \ion{Mg}{ii}, \ion{Al}{i}, \ion{Al}{ii}, \ion{Si}{i}, \ion{Si}{ii}, \ion{Ca}{i}, \ion{Ca}{ii}, \ion{Fe}{i}, \ion{Fe}{ii}, an approximation for the remaining metals, and CH, OH), free-free transitions for \ion{H}{i}, H$^-$, \ion{He}{i}, He$^-$, \ion{C}{i}, \ion{C}{ii}, C$^-$, N$^-$, O$^-$, \ion{Mg}{i}, \ion{Si}{i}, H$_2^+$, H$_2^-$, CO$^-$, and H$_2$O$^-$, collision induced absorption for \ion{H}{i}+\ion{H}{i}, \ion{H}{i}+\ion{He}{i}, H$_2$+\ion{H}{i}, H$_2$+H$_2$, H$_2$+\ion{He}{i}, scattering by electrons, as well as Rayleigh scattering by \ion{H}{i}, H$_2$, and \ion{He}{i}. References to all the data used are provided in \cite{Gustafsson2008}.
The H$^-$ opacity was improved in the present version from the original \cite{Wishart1979} to the more recent \cite{Mclaughlin2017b,Mclaughlin2017a}. For the solar model, the difference in the optical and near infrared is less than $0.5\%$.
Bound-free opacity for NH was also added following \citet{Shen2014} (P.C. Stancil, priv. comm.).
Hydrogen bound-free and line opacity is treated using the code of \cite{Barklem2015}. Line broadening is included through a Voigt profile resulting from the convolution of the Gaussian microturbulence and thermal broadening profile with the natural and collisional broadening Lorentz profile. Broadening caused by elastic hydrogen collisions is treated with the so-called 'ABO' (Anstee, Barklem, O'Mara) theory \citep{Anstee1995, Barklem1997,Barklem2000}. Electron (linear Stark) collisional broadening has also been added in the present version.

The NLTE version of TS uses grids of NLTE departure coefficients to compute NLTE line profiles. This is done by correcting the line opacity and the line source functions of all lines, similar to other spectrum synthesis and abundance analyses codes, such as SIU and SME. The NLTE extinction coefficient is obtained from the LTE value $\alpha_\lambda^*$ using:
\begin{equation}
    \alpha_\lambda = \alpha_\lambda^* b_l \frac{1-\frac{b_u}{b_l}e^{-h\nu/kT}}{1-e^{-h\nu/kT}},\label{eq-bb}
\end{equation}
where $b_u$ and $b_l$ are the departure coefficients for the upper and lower levels, respectively, the other symbols having their usual meaning.
The emissivity of a line is given by
\begin{equation}
j_\lambda = \alpha_\lambda B_\lambda = \alpha_\lambda^* b_u B_\lambda, \label{eq-jbb}
\end{equation}
where $B_\lambda$ is the Planck function,
and the source function at a given wavelength is obtained by summing the emissivity from the continuum and all contributing lines:
\begin{equation}
S_\lambda=\frac{\sum{j_\lambda}}{\sum{\alpha_\lambda}}\label{eq-Sbb}.
\end{equation}

For the elements that are computed in LTE, the standard Saha-Boltzmann distribution functions are used.

TS can compute large chunks of spectra, including lines from all species, some elements in LTE, others in NLTE, allowing the treatment of blends. The elements to be calculated in NLTE are handled by the user via a dedicated input file. A new feature allows to specify a list of spectral windows to be calculated, reducing computing time and the size of the output. There is no provision for taking into account departures from LTE for the continuum opacities yet, but we do not expect any significant differences associated with this approximation in the optical and near-IR wavelength ranges, which are typically used as diagnostics in stellar spectroscopy. \citet{Haberreiter2008} show that using NLTE opacity distribution functions in the calculations of the solar model atmosphere has a significant effect only at wavelengths shorter than 260~nm (UV). \citet{Young2014} investigated NLTE model atmospheres for red giants. Also their work shows that NLTE effects influence the spectral energy distribution in stars with $\teff \gtrsim 4000$~K only below $400$~nm.
\subsection{Line list}\label{sec:linelist}
Concerning atomic and molecular data, we make use of the Gaia-ESO line list \citep{Heiter2021}, which covers the wavelength ranges
from 475 to 685~nm and from 850 to 895~nm.
The atomic part of the line list is mainly based on experimental data supplemented by the most reliable theoretical data (see \citealt{Heiter2021} for a detailed description of the data sources).
For the wavelength ranges not covered by the Gaia-ESO line list, we use data from the VALD database\footnote{\url{http://vald.astro.uu.se}} \citep{Pisk:95,2015PhyS...90e4005R}. 
It is important to note that the line list was not calibrated on any observational data, which is sometimes done in the literature (e.g., \citealt{2021AJ....161..254S}) and therefore it is not tied to any particular type of star or stellar model, which makes it universally applicable to any object in the entire parameter space of cool stars.
\subsection{Statistical equilibrium}\label{sec:multi}
The grids of departure coefficients were computed using the 1D NLTE radiative transfer solver MULTI \citep{Carlsson1986} updated as described in \citet{Bergemann2019} and \citet{Gallagher2020}. Statistical equilibrium of a NLTE element in MULTI is computed under the standard assumption of a trace element, that is assuming that deviations from LTE do not influence the structure of the input model atmosphere. 

The departure grids were computed for $13$ chemical elements, namely H, O, Na, Mg, Si, Ca, Ti, Mn, Fe, Co, Ni, Sr, and Ba. The NLTE model atoms were taken from the following studies:
\begin{itemize}
    \item H: \citet[][]{Mashonkina2008}
    \item O: \citet[][]{Bergemann2021}
    \item Na: \citet[][]{Larsen2021},
    \item Mg: \citet[][]{Bergemann2017}
    \item Si: \citet[][]{Bergemann2013} with Si$+$H collisional data from \citet{Belyaev2014} and updates to the radiative data as described in 
    \citet{Magg2022}
    \item Ca: \citet[][]{Mashonkina2007}, updates described in \citet{Semenova2020}
    \item Ti: \citet[][]{Bergemann2011} 
    \item Mn: \citet[][]{Bergemann2019}
    \item Fe: \citet[][]{Bergemann2012}, updates described in \citet{Semenova2020}
    \item Co: \citet[][]{Bergemann2010} with Co$+$H collisional data from \citet{Yakovleva2020}
    \item Ni: \citet[][]{Bergemann2021} with Ni$+$H collisional data from \citet{Voronov2022}
    \item Sr: \citet[][]{Bergemann2012c} with Sr$+$H collisional data from Gallagher et al. in prep
    \item Ba: \citet[][]{Gallagher2020}
\end{itemize}

For Fe, we solved the radiative transfer and statistical equilibrium equations for each metallicity ([Fe/H]) point in the MARCS and STAGGER model grids. For all other chemical elements, we solved the same equations varying the element abundance within [$-$2, +1] dex in abundance steps of $0.1$ dex relative to the corresponding solar ratio [X/Fe] (is always zero by definition). The grid of NLTE departure coefficients is available for each of the 12 chemical elements, described above.\footnote{The current grids of departure coefficients can be found at \url{https://keeper.mpdl.mpg.de/d/9b6c265057aa4bceb939/}.}

The NLTE grids of departures were computed for each model atmosphere in the MARCS and STAGGER grids (as described in Sect. 2.1) and are supplied with the public release of the code. We recommend to use the model atmosphere grids supplied with the grids of departure coefficients, because we also provide a dedicated interpolating function (based on the Fortran code by Thomas Masseron, see also \citealt[][]{Gustafsson2008}) that takes the rectangular grids of model atmosphere and the corresponding grids of departures coefficients and produces an interpolated atmosphere structure and departure file for any desired combination of stellar parameters. We adjusted this module to also interpolate the grid of departure coefficients in $4$ dimensions: $\teff$, $\logg$, $\feh$, and elemental abundances. We also have created a new module that provides the same function for model atmospheres formatted for use with MULTI to handle the interpolation of the $\td$ models from STAGGER.\footnote{The code to simultaneously interpolate model atmospheres and departure coefficient grids is included on the TS Github.}
\begin{figure*}[ht!]
\centering
\includegraphics[width=0.9\linewidth]{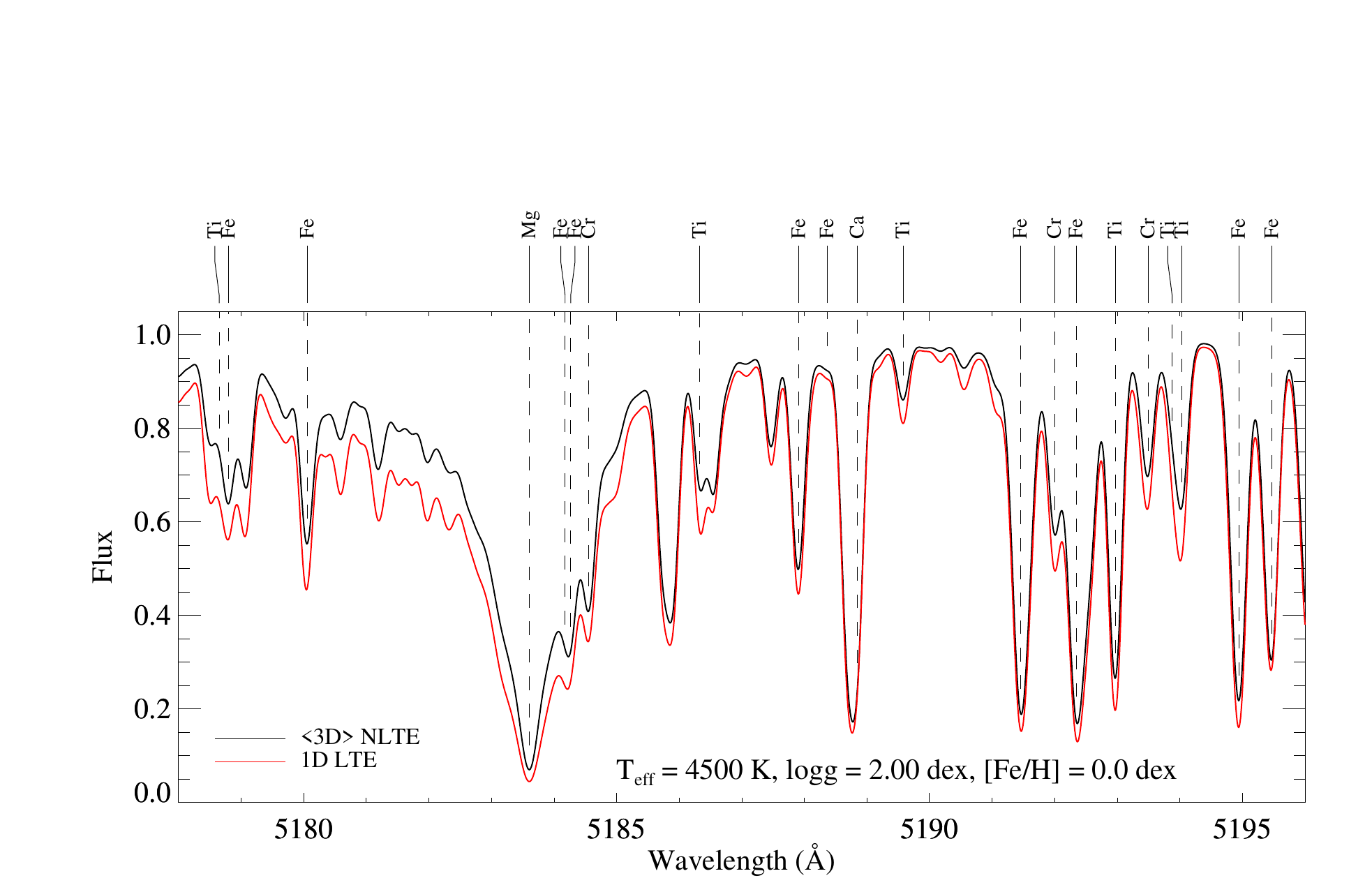}
\caption{Synthetic spectrum computed with a <3D> Stagger model atmosphere in NLTE, compared with a spectrum calculated in 1D LTE. The stellar parameters used in the calculations are indicated in the figure inset.}
\label{fig:TurbTauMARCSvsSTAGGER}
\end{figure*}

\section{New capabilities of Turbospectrum}\label{sec:newturbo}

\subsection{Comparison with NLTE calculations from MULTI}\label{sec:tests}

We tested our method for calculating NLTE line profiles using TS by generating NLTE model spectra using the MULTI code, and by comparing the equivalent widths (EWs) of various Ca lines. MULTI is the code that we use to calculate the NLTE statistical equilibrium and NLTE grids of departures, and so it provides the most self-consistent anchor point for our TS results. The same four sets of atmospheric parameters as in Fig. \ref{fig:TeffTauMARCSvsSTAGGER} were taken from grid points of the MARCS model atmosphere grid and chosen as test cases to represent giant and dwarf stars at solar and low metallicity. As mentioned previously, the giant stars were given an effective temperature of 4000~K and a surface gravity of $\logg = 2.0$, and the dwarfs were given an effective temperature of 6000~K and a surface gravity of $\logg = 4.5$. Our low metallicity model used an [Fe/H] of $-2.0$ dex.

The results of this test for each of the model spectra are shown in Table \ref{table:ca-percentdiff}. Direct comparisons of NLTE EWs are not useful, because small discrepancies may arise due to other differences between the two codes in, e.g., equation of state, background opacities, or radiative transfer solvers and angle quadrature. We therefore calculated with each code the ratio of NLTE to LTE EWs for a given set of Ca lines. Ca was chosen as the element, for which the NLTE abundance corrections show a large dynamic range, being positive for some lines and negative for the others. The average percent difference of the ratio of EWs ($EW_{NLTE}/EW_{LTE}$) computed with MULTI and TS is given in the final column of the table. All lines show a difference below $0.5\%$, with the average being below $0.1\%$, showing that the TS predicts correct NLTE effects in spectral lines.

We do not detect any systematic bias against the atomic properties of the transition, the lower and upper level excitation potential, oscillator strength, or wavelength. Also there is no evidence that either of the codes systematically over- or under-estimates the line EWs. A similar test was performed for Fe lines, confirming our results for Ca. Therefore, we conclude that our method for calculating NLTE line profiles with TS is robust.
\begin{table}
\caption{The average percent difference between TS and MULTI calculation of the ratio of NLTE to LTE equivalent width of nine Ca lines computed for four different sets of parameters. For all results, MARCS model atmospheres were used. For the following table, $R_{NLTE}$ = $EW_{NLTE}/EW_{LTE}$.}
\label{table:ca-percentdiff}
\centering
\begin{tabular}{c c c c}
\hline\hline
 $\teff$ & $\logg$ & [Fe/H] &  $R_{NLTE}$ (TS) - $R_{NLTE}$ (MULTI) (in \%)\\
\hline
6000 & 4.5 & 0.0    & 0.10 \\
6000 & 4.5 & $-2.0$ & 0.10 \\
4500 & 2.0 & 0.0    & 0.03 \\
4500 & 2.0 & $-2.0$ & 0.05\\
\hline
\hline
\end{tabular}
\end{table}
\subsection{NLTE calculations with multiple elements} \label{sec:nltemultiple}

The main novelty  of the new TS is that it can compute  NLTE spectra, treating multiple elements in NLTE at the same time. Computing multiple elements in NLTE is crucial for studies that want to consider blended lines for abundance determination or study large regions of spectra where NLTE contributions to multiple elements are important.

To test and showcase this capability, we generated spectra with $\td$ NLTE line profiles for H, O, Mg, Ca, Mn, and Fe  for all four test cases that we used to compare our NLTE calculations with MULTI, and compared them to spectra generated using 1D LTE. The model spectra were then convolved to a resolution of $R = \lambda/\Delta\lambda \sim$ 20\,000 to demonstrate how the differences between the two models would appear at a resolution similar to those used for future surveys such as WEAVE and 4MOST. Figure \ref{fig:TurbTauMARCSvsSTAGGER} shows these two spectra for comparison in a window centered on 5190~\AA. There are several areas where the 1D LTE model differs significantly from the $\td$ NLTE model.
%

\subsection{H line profiles} \label{sec:hprofiles}

The new version of TS also has the ability to calculate H line profiles using NLTE. The wings of \ha\ are arguably the most sensitive and powerful spectroscopic diagnostics of effective temperature for dwarf stars \citep[e.g.][]{Fuhrmann1993,Fuhrmann1994,Grupp2004a,Grupp2004b}. It is therefore important to compare the H line profiles generated with NLTE physics with LTE profiles and observations. 

\begin{figure*}[h!]
    \centering
    \includegraphics[width=\linewidth]{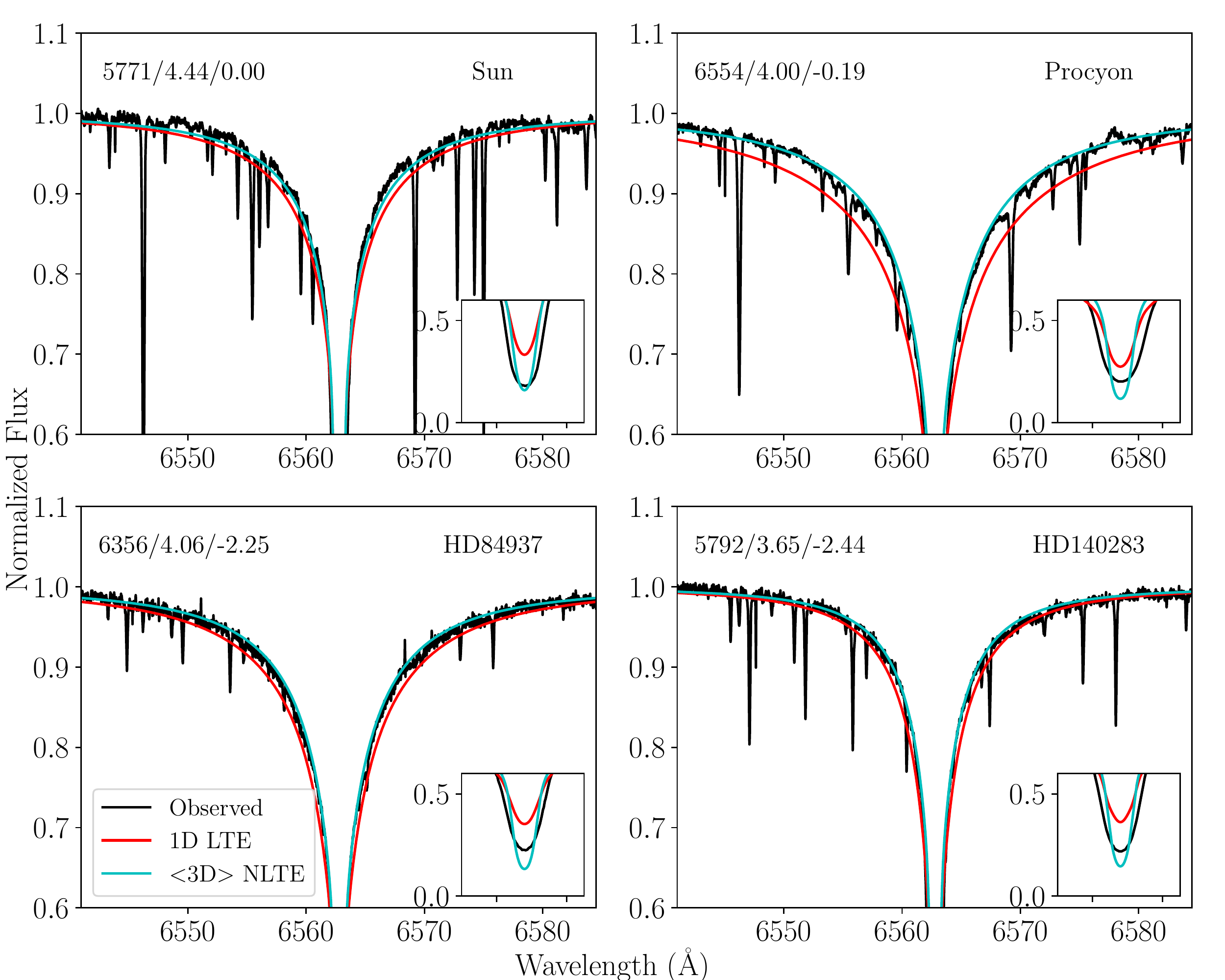}
    \caption{The \ha\ line profiles of the Sun, Procyon, HD~84937, and HD~140283. The black lines are observations made by UVES. The red and cyan lines are the 1D LTE and $\td$ NLTE model profiles, respectively, generated with Turbospectrum. Inset in the lower right corner of each panel is a zoomed in look at the core of each line.}
    \label{fig:hlines}
\end{figure*}

Figure \ref{fig:hlines} shows the simulated $\td$ NLTE and 1D LTE \ha\ lines, respectively, compared to the observed spectra for the Sun, Procyon, HD~84937, and HD~140283.  The model lines were calculated using independent stellar parameters for these targets, obtained using non-spectroscopic diagnostics (see Sect.~\ref{sec:observed}). 

Clearly, the $\td$ NLTE models provide an excellent fit to the \ha\ wings that supports previous studies of H lines in the literature \citep{Amarsi2018}. In contrast, the 1D LTE models fail to reproduce the wings, leading to a systematic under-estimation when using the line to derive $\teff$. The line cores cannot be described even in $\td$ NLTE. This is not a concern in this work, because it is well known that the Balmer line cores are formed in chromospheres, and so require magneto-hydrodynamical modelling and Non-LTE for a self-consistent description \citep{Leenaarts2012,Bergemann2016}.

\section{Example abundance analysis with NLTE Turbospectrum}\label{sec:exampleAnalysis}
\subsection{Observational data}\label{sec:observed}

Our observational sample is drawn from the Gaia FGK benchmark star sample \citep{Jofre2018}, and we refer the readers to \citet{Jofre2014} and \citet{Heiter2015} for the details on the methods used for the determination of the atmospheric parameters and abundances. In short, the sample comprises 31 FGK-type stars covering a broad range of $\teff$, $\logg$, and $\feh$ and it includes main-sequence dwarfs, turn-off stars, subgiants, and red giants. For all of these stars, spectra collected with the UVES and HARPS spectrographs at the ESO Very Large Telescope (VLT) and the La Silla 3.6m telescope are available and we make use of these data in this work. The typical signal-to-noise ratio of the data exceeds $1\,000$, and the resolving power is $R=47\,000$ for UVES and $R=115\,000$ for the HARPS data. The spectra cover a broad wavelength range of 480 to 680 nm (UVES) and 378 to 691~nm (HARPS). For more details on the spectra and their reduction, including the continuum normalisation, we refer the reader to \citet{Blanco-Cuaresma2014}.

We adopt the reference stellar parameters for these stars from \citet{Jofre2018}, with a few updates as described in \citet{Gent2022}. The $\teff$ estimates for the majority of the stars are based on angular diameters determined from interferometric data collected with several different facilities, such as CHARA or the VLTI, together with bolometric flux estimates. For a few stars the angular diameters were determined from surface-brightness relations using ($B-K$) and ($V-K$) colour indices or from infrared photometry \citep[][Sect.~3]{Heiter2015}.
The surface gravities were determined using angular diameters, Hipparcos parallaxes\footnote{The Hipparcos parallaxes are consistent with Gaia DR2 and Gaia eDR3 values for these stars.}, and masses estimated from evolutionary tracks \citep[][Sect.~4]{Heiter2015}. For the stars in the sample which have asteroseismic data it was verified that the surface gravities determined from the angular diameters agree with those calculated from the frequency of maximum power and the $\teff$ values \citep[][Sect.~5.3]{Heiter2015}.
Metallicities and microturbulence estimates were determined by fitting Fe line profiles or equivalent widths in LTE using MARCS models and several different radiative transfer codes and applying NLTE corrections to the resulting Fe line abundances \citep{Jofre2014}. For some of the stars (Procyon, HD 122563, HD 140283) more accurate metallicities computed using a full 3D NLTE approach became available recently \citep{Amarsi2016}. Therefore, we also use 3D NLTE [Fe/H] values, where possible.

\subsection{Fitting stellar parameters with Turbospectrum}\label{sec:tssapp}
\begin{figure*}[ht!]
    \centering
    \includegraphics[width=\linewidth]{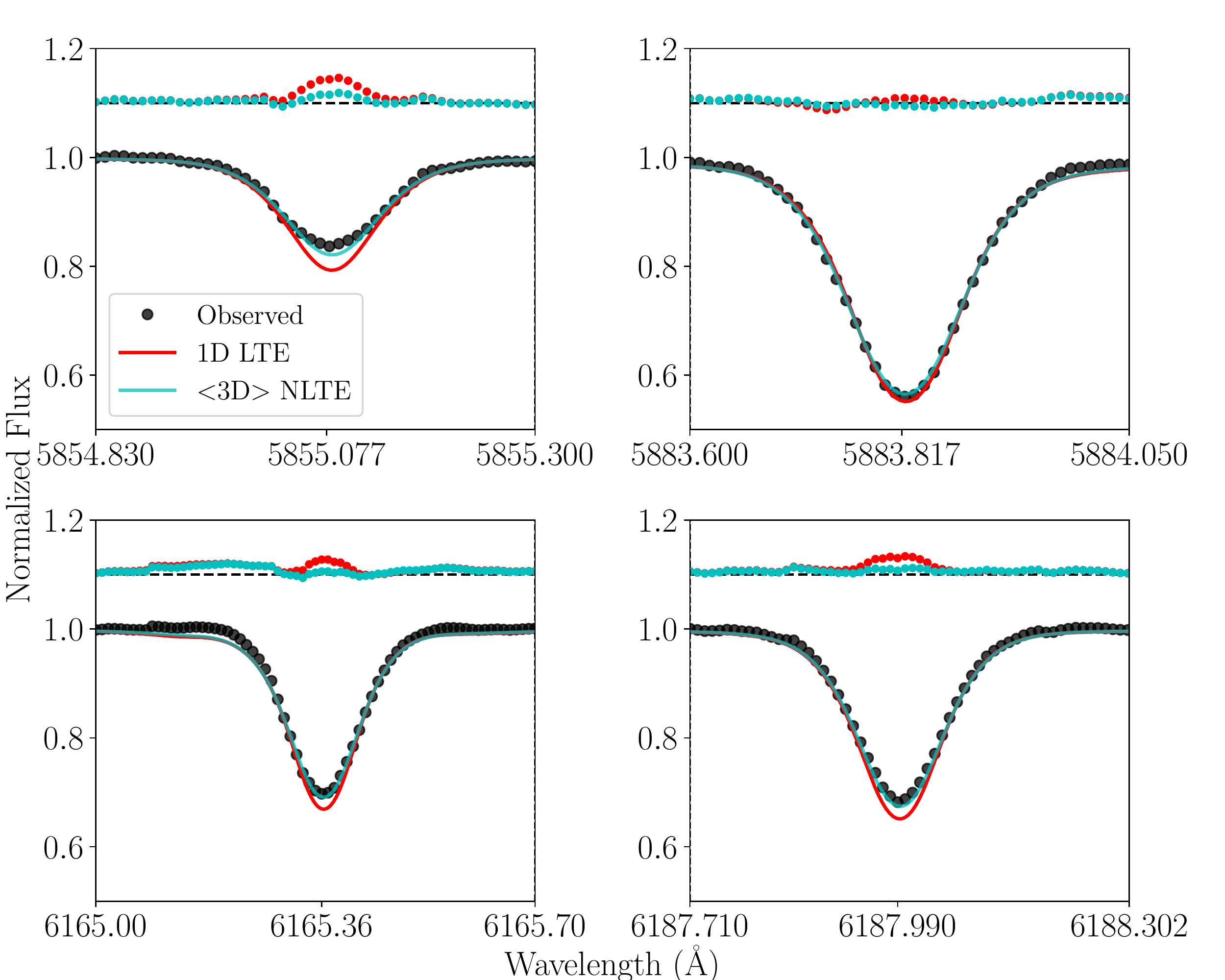}
    \caption{Normalized flux vs. wavelength for four sample \ion{Fe}{i} lines in an observed solar spectrum together with model fits. Observations are shown as black circles, 1D LTE model fits are shown as red lines, and $\td$ NLTE model fits are shown as cyan lines. 
    Residuals for the 1D LTE and $\td$ NLTE model fits are shown as red and cyan points, respectively. The points have been shifted by 0.1 and are plotted above each spectrum with a dashed line to indicate the zero point at 1.1.}
    \label{fig:sun-fe-lines}
\end{figure*}

For this work, we developed a Python module that wraps NLTE TS, which can be used to directly determine stellar parameters, including $\teff$, $\logg$, $\feh$, $\vmic$, $\vmac$, abundances, etc, from the observed stellar spectrum. That is, TS can be used to not only generate synthetic grids, but also as a standalone code to perform spectrum synthesis and stellar parameter diagnostics. This method allows us to instantly update the macro- and microphysics (atomic and molecular data) of our models and seamlessly switch between 1D NLTE, 1D LTE, $\td$ LTE, or $\td$ NLTE between runs without any overhead. This also allows us to directly investigate the quality of the spectral line fits and enhances the flexibility of the analysis, as the abundance of any chemical element in the periodic table can be determined directly, provided a suitable line list exists.
\begin{figure*}[ht!]
    \centering
    \includegraphics[width=\linewidth]{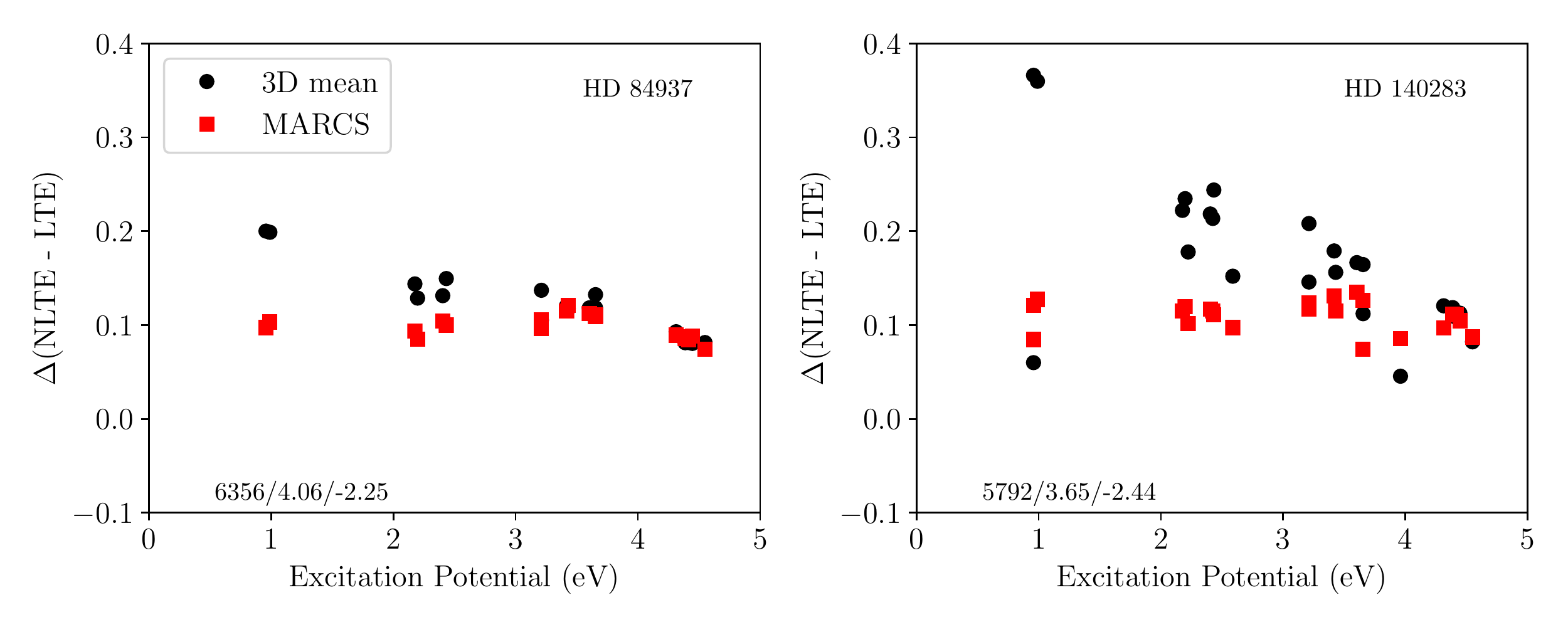}
    \caption{Each panel shows the difference between the line abundances derived from a NLTE and an LTE fit for both $\td$ (Stagger) and MARCS model atmospheres for various \ion{Fe}{i} lines for HD~84937 and HD~140283 as a function of excitation potential of the line. Fits using MARCS model atmospheres are shown as red squares and those using $\td$ model atmospheres are shown as black circles. The differences are given in units of dex.}
    \label{fig:diff-fe-vs-excitpotent}
\end{figure*}

We use the Nelder-Mead numerical method (also known as the simplex method) to fit a set of observed lines within a pre-defined spectral range with a model spectrum. The line choice is controlled by a set of user-defined line and continuum masks. The Nelder-Mead method uses a simplex to test $n + 1$ points in an $n$-dimensional space until the minimum of a function consisting of $n$ dimensions is found. We apply this method to a given set of spectral windows centered on spectral features of the element under consideration. The element abundance, a radial velocity shift, and - if needed - other parameters (like microturbulence) are simultaneously adjusted until the $\chi^2$ value is minimized and a solution to both parameters is determined. The selection of diagnostic lines is described in Sect. \ref{sec:linelist}.

For fitting multiple lines at a time in a single spectrum, we find that the precision of our abundances is improved by setting the microturbulence based on an empirical relation related to the stellar parameters of the star being fit \citep[see][and references therein for a description of the relation used]{Buder2021}. We also adopt a macroturbulence value of 3.5 km/s for all stars in the sample and apply a broadening correction to the synthetic spectra to get line profiles that better match the observed spectra. Our Python module used to conduct the fitting handles the convolution of synthetic spectra automatically.

In Fig. \ref{fig:sun-fe-lines}, we show <3D> NLTE and 1D LTE profiles of selected \ion{Fe}{i} lines. Both models use the same input parameters and input Fe abundance, the fit being optimized for the NLTE case. The quality of the fit is very good, suggesting that the new approach can be used for a robust spectrum synthesis in the parameter space described by our input models.

Our Python wrapper used for fitting can be found at the following Github page \url{https://github.com/JGerbs13/TSFitPy}. It makes use of the model interpolator mentioned in Section \ref{sec:multi} and the latest version of TS, both of which can be found on the main TS Github page. In addition to our fitting module, we also include a script that can be used to run the model interpolator and generate a synthetic spectrum with TS for a given set of stellar parameters.
\section{Abundance results} \label{sec:abundResults}

\subsection{Metallicity}

Figure \ref{fig:diff-fe-vs-excitpotent} shows the differences between NLTE and LTE abundances of \ion{Fe}{i} lines for two example stars in the sample. The atomic parameters of \ion{Fe}{i} lines used for these fits are given in Table~\ref{table:Fe_lines}. The abundances were computed using 1D LTE, 1D NLTE, $\td$ LTE, and $\td$ NLTE. These figures confirm that the 1D LTE and $\td$ LTE models generally under-estimate the abundances derived from the lines of neutral species, similar to other findings in the literature \citep{Bergemann2012, Amarsi2016, Sitnova2016, Lind2017,Bergemann2019}. We find no trend with the line EW, other than the fact that weak lines (especially those with EW $< 5$~mA) show a higher spread in abundance for the same star than strong lines, as expected. Similarly, there is no clear trend in abundance with lines arising from different energy levels. However, there does appear to be a trend in the difference between NLTE and LTE abundance with the excitation potential for $\td$ models. As seen in Figure \ref{fig:diff-fe-vs-excitpotent}, lines with higher excitation potential tend to have a lower $\Delta$(NLTE $-$ LTE). This is consistent with the results by \citet{Bergemann2012} and \citet{Amarsi2016}, who found the same trends in their analyses of Fe lines in FGK-type stars. 

\begin{figure}[h!]
    \centering
    \includegraphics[width=\linewidth]{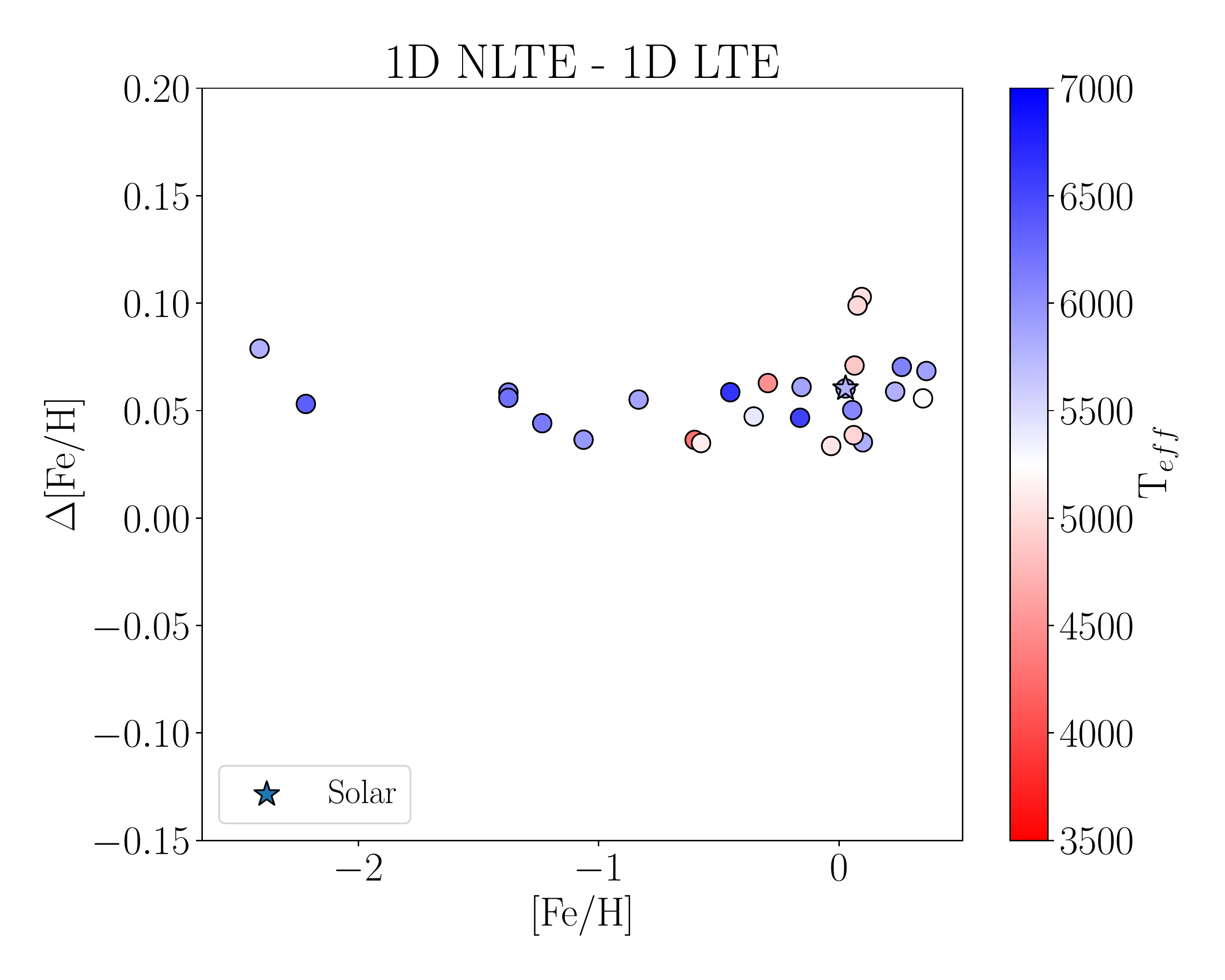}
    \includegraphics[width=\linewidth]{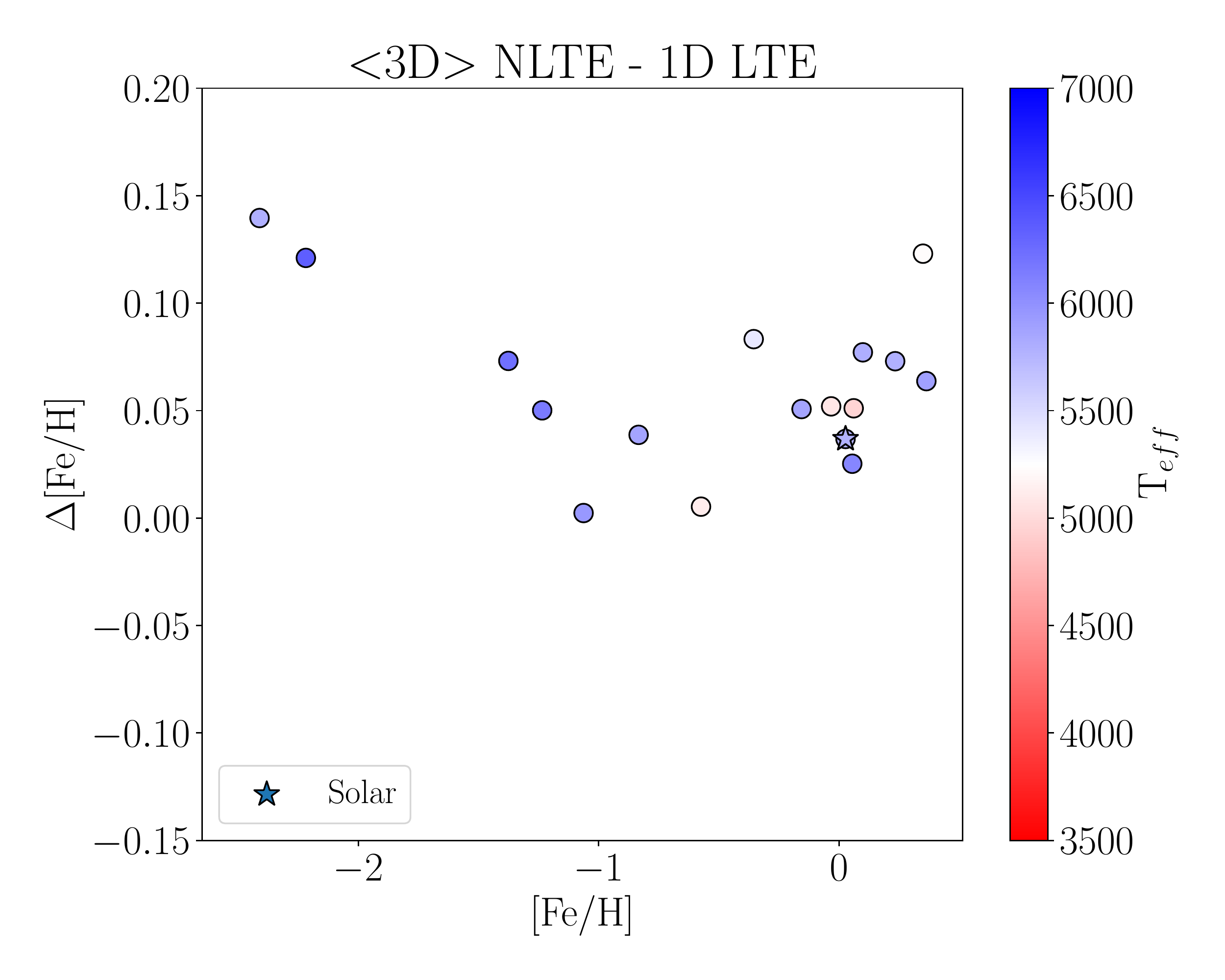}
    \caption{Difference in the [Fe/H] value determined using 1D NLTE (top) and $\td$ NLTE (bottom) and 1D LTE, as a function of [Fe/H]. Symbols are color coded based on the effective temperature of the star. Fits to spectra of the Sun are indicated as stars.}
    \label{fig:1dnlte-lte}
\end{figure}

Figure \ref{fig:1dnlte-lte} shows the differences between 1D NLTE and $\td$ NLTE abundances and the 1D LTE results. We note that the available $\td$ model atmosphere grid is limited in terms of the $\teff$-$\logg$ coverage, therefore, in the attempt to avoid extrapolation, we could not determine Fe abundances for selected types of stars (very cool red giants and red supergiants). The differences are systematic, which simply reflects the smooth and regular behaviour of NLTE effects with the $\teff$, $\logg$, and $\feh$ of the star. As shown in previous work \citep{Bergemann2012, Bergemann2014}, NLTE effects on \ion{Fe}{i} lines grow with increasing $\teff$, and decreasing surface gravity and metallicity. In hotter or more metal-poor stars, the more intense UV radiation field enhances over-ionisation, which is the main driver of NLTE effects in neutral species with large photo-ionisation cross-sections and low ionisation potentials. In lower $\logg$ stars (lower densities), larger NLTE effects are rather caused by less efficient collisional thermalisation. For most stars we find an average difference of about $+0.05$~dex between the [Fe/H] determined using 1D NLTE and 1D LTE. 
When $\td$ model atmospheres are used (Figure \ref{fig:1dnlte-lte}, bottom panel), the differences for the stars in our sample fall between $0.00$ and $+0.15$ dex with lower metallicity stars showing the largest differences. This value for the overall difference and slight trend with metallicity agrees well with differences expected by model calculations \citep{Bergemann2012, Amarsi2016, Semenova2020}. 

\begin{figure}[h!]
    \centering
    \includegraphics[width=\linewidth]{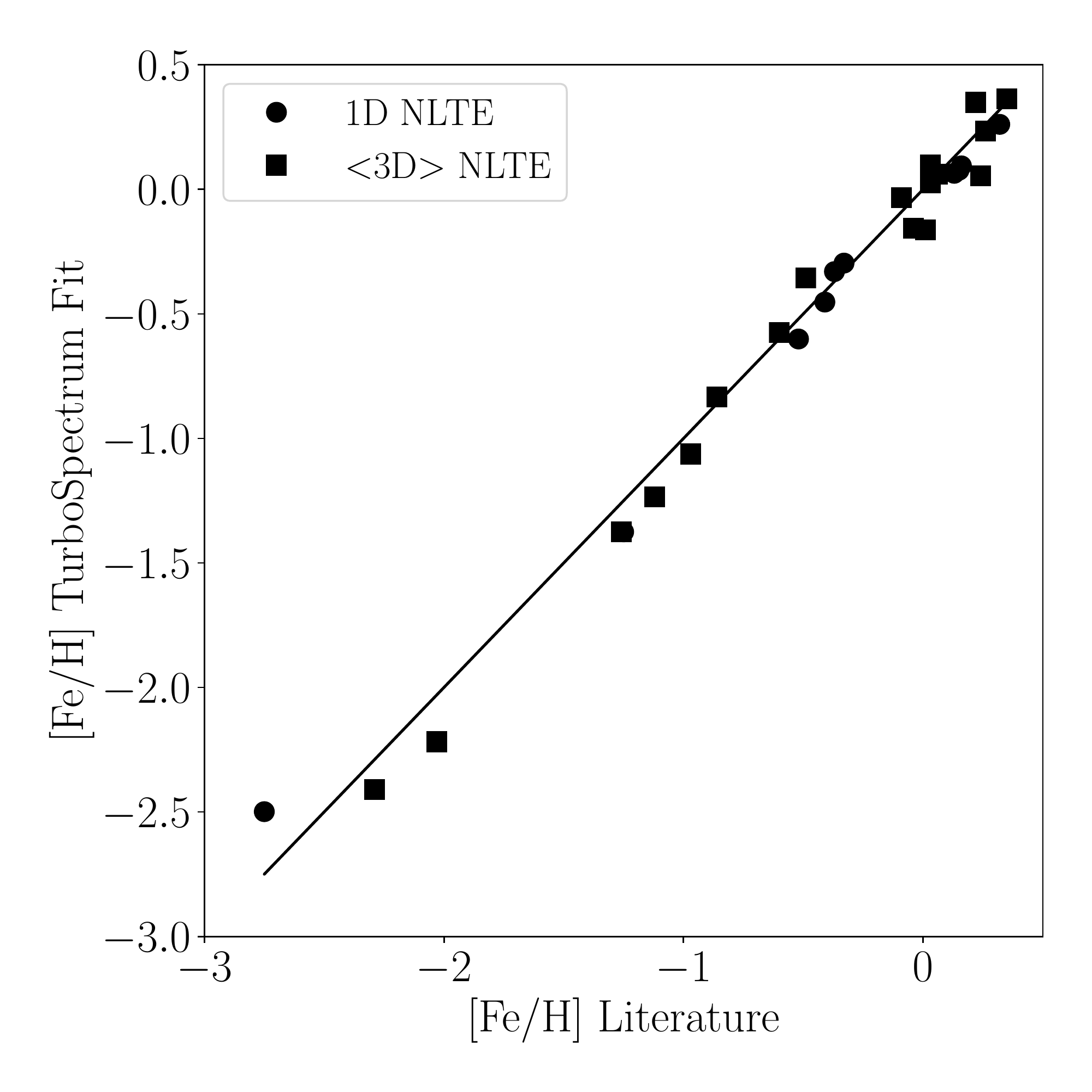}
    \caption{[Fe/H] determined from the Turbospectrum fit versus values from the literature (see text). For the TS fit the adopted value is the <3D> NLTE fit, or the 1D NLTE in cases where a <3D> NLTE fit was not available.}
    \label{fig:fe-lit-compare}
\end{figure}

Figure \ref{fig:fe-lit-compare} shows the final $\td$ NLTE or 1D NLTE metallicities determined by minimizing the total $\chi^2$ of all Fe lines. The values are compared to the 1D NLTE values from \citet{Jofre2018}. Generally, we find that the metallicities are in good agreement with those in the literature within the uncertainties, with no systematic offsets between stars observed with either instrument (HARPS or UVES). We emphasize, however, that there is no direct model-independent method to derive abundances, which is why accuracy cannot be tested this way. Our approach, using $\td$ model atmospheres, is physically more realistic compared to \citet{Jofre2018}, therefore differences in results are expected. For the Sun, we obtain a $\td$ NLTE iron abundance of $\log A(\rm Fe) = 7.50 $, which is consistent within the uncertainty with the 3D NLTE estimate by \citet{Lind2017} and \citet{Magg2022}. Minor differences are expected not only because of methodological differences ($\td$ NLTE vs 3D NLTE, different line selection), but also because solar intensities, rather than fluxes, were used in \citet{Lind2017}. Our 1D LTE, 1D NLTE, and $\td$ NLTE results are given in Table \ref{table:Fe}. 

\begin{table*}
\caption{\ion{Fe}{i} lines used for fitting.}
\label{table:Fe_lines}
\centering
\begin{tabular}{c c r c r}
\noalign{\smallskip}\hline\noalign{\smallskip}
$\lambda$ & $E_{\rm low}$ & $\log gf$ & vdW\tablefootmark{$a$} 
& Ref.\tablefootmark{$b$}
\\
(\AA) & (eV) & & & \\
\noalign{\smallskip}\hline\noalign{\smallskip}
5141.739 & 2.424 & $-$1.978 &  367.251 &  1 \\ 
5217.389 & 3.211 & $-$1.074 &  815.232 &  2 \\ 
5242.491 & 3.634 & $-$0.967 &  361.248 &  3 \\ 
5324.179 & 3.211 & $-$0.108 &  791.236 &  4 \\ 
5364.871 & 4.446 &    0.228 & 1013.281 &  3 \\ 
5365.399 & 3.573 & $-$1.020 &  283.261 &  3 \\ 
5367.466 & 4.415 &    0.444 &  972.280 &  5 \\ 
5379.574 & 3.695 & $-$1.514 &  363.249 &  3 \\ 
5383.369 & 4.313 &    0.645 &  836.278 &  3 \\ 
5398.279 & 4.446 & $-$0.630 &  993.280 &  6 \\ 
5415.199 & 4.387 &    0.643 &  910.279 &  3 \\ 
5445.042 & 4.387 & $-$0.020 &  895.279 &  7 \\ 
5501.465 & 0.958 & $-$3.046 &  239.249 &  3 \\ 
5506.779 & 0.990 & $-$2.795 &  241.248 &  8 \\ 
5543.936 & 4.218 & $-$1.040 &  742.238 &  6 \\ 
5560.211 & 4.435 & $-$1.090 &  895.278 &  6 \\ 
5569.618 & 3.417 & $-$0.517 &  848.233 &  9 \\ 
5576.089 & 3.430 & $-$0.900 &  854.232 &  6 \\ 
5638.262 & 4.220 & $-$0.720 &  730.235 & 10 \\ 
5661.345 & 4.284 & $-$1.765 &  765.209 & 11 \\ 
5679.023 & 4.652 & $-$0.820 & 1106.291 &  6 \\ 
5731.762 & 4.256 & $-$1.200 &  727.232 &  6 \\ 
5741.848 & 4.256 & $-$1.672 &  725.232 &  3 \\ 
5855.076 & 4.608 & $-$1.478 &  962.279 & 11 \\ 
5883.816 & 3.960 & $-$1.260 &  998.250 &  6 \\ 
5905.671 & 4.652 & $-$0.690 &  994.282 &  6 \\ 
5930.180 & 4.652 & $-$0.230 &  983.281 & 12 \\ 
5956.694 & 0.859 & $-$4.599 &  227.252 & 13 \\ 
6024.058 & 4.549 & $-$0.120 &  823.275 & 12 \\ 
6027.051 & 4.076 & $-$1.089 &  380.250 &  3 \\ 
6056.004 & 4.733 & $-$0.320 & 1029.286 & 10 \\ 
6093.643 & 4.608 & $-$1.400 &  866.274 &  6 \\ 
6151.617 & 2.176 & $-$3.295 &  277.263 & 14 \\ 
6165.360 & 4.143 & $-$1.473 &  380.250 &  3 \\ 
6173.334 & 2.223 & $-$2.880 &  281.266 & 15 \\ 
6187.989 & 3.943 & $-$1.620 &  903.244 &  6 \\ 
6219.280 & 2.198 & $-$2.432 &  278.264 & 14 \\ 
6246.318 & 3.603 & $-$0.771 &  820.246 &  2 \\ 
6252.555 & 2.404 & $-$1.699 &  326.245 & 16 \\ 
6265.132 & 2.176 & $-$2.550 &  274.261 & 16 \\ 
6270.224 & 2.858 & $-$2.470 &  350.249 &  1 \\ 
6297.792 & 2.223 & $-$2.737 &  278.264 & 14 \\ 
6301.500 & 3.654 & $-$0.720 &  832.243 & 17 \\ 
6322.685 & 2.588 & $-$2.430 &  345.238 & 18 \\ 
6330.848 & 4.733 & $-$1.640 &  915.277 &  6 \\ 
6335.330 & 2.198 & $-$2.177 &  275.261 &  3 \\ 
6336.823 & 3.686 & $-$0.852 &  845.240 &  9 \\ 
6393.601 & 2.433 & $-$1.452 &  326.246 &  1 \\ 
6411.648 & 3.654 & $-$0.596 &  820.247 &  2 \\ 
6430.845 & 2.176 & $-$2.005 &  271.257 & 16 \\ 
6481.870 & 2.279 & $-$2.981 &  308.243 & 14 \\ 
6498.938 & 0.958 & $-$4.687 &  226.253 & 13 \\ 
\\
\hline\noalign{\smallskip}
\end{tabular}
\tablefoot{
\tablefoottext{$a$}{Van der Waals broadening parameter from \citet{Barklem2000}.}
\tablefoottext{$b$}{References for $gf$-values:
(1) average of \citet{BKK} and \citet{BWL}
(2) average of \citet{BKK}, \citet{BWL}, and \citet{GESHRL14}
(3) \citet{BWL}
(4) average of \citet{BKK} and \citet{2014MNRAS.441.3127R}
(5) average of \citet{BK} and \citet{BWL}
(6) \citet{MRW}
(7) \citet{1970ApJ...162.1037W} renormalized to \citet{FMW}
(8) average of \citet{GESB79b} and \citet{BWL}
(9) average of \citet{BK} and \citet{GESHRL14}
(10) \citet{2014MNRAS.441.3127R}
(11) \citet{BK}
(12) \citet{WBW} renormalized to \citet{FMW}
(13) average of \citet{GESB86} and \citet{BWL}
(14) average of \citet{BKK}, \citet{GESB82c}, and \citet{BWL}
(15) \citet{GESB82c}
(16) average of \citet{GESB82d} and \citet{BWL}
(17) average of \citet{BKK} and \citet{BWL}
(18) average of \citet{GESB82c} and \citet{BWL}
}
}
\end{table*}

\renewcommand{\footnoterule}{} 
\begin{table*}
\begin{minipage}{\linewidth}
\caption{Results for \ion{Fe}{i} ([Fe/H] using a solar abundance of $\log A(\rm Fe) = 7.50$).}
\label{table:Fe}
\centering
\begin{tabular}{c c c c c c c c}
\hline\hline
Star name & Adopted $\rightarrow$ & & Literature\footnote{The stellar parameters come from \citet{Jofre2018}, with a few updates as described in \citet{Gent2022}. We also include the literature [Fe/H] value from \citet{Jofre2018} that was derived using 1D NLTE as a reference.}  & Derived $\rightarrow$  \\
& $\teff$ [K] & $\logg$ [cms$^{-2}$] & [Fe/H] & 1D LTE & 1D NLTE  & $\td$ LTE  & $\td$ NLTE  \\
\hline
    18Sco &  5810 &  4.44 & 0.03 &   -0.01 &     0.03 &          0.02 &           0.07 \\
 Arcturus &  4286 &  1.60 & -0.52 &  -0.67 &    -0.63 &        ... &         ... \\
    $\mu$ Ara &  5902 &  4.30 &  0.35 &   0.27 &     0.34 &          0.31 &           0.33 \\
   $\eta$ Boo &  6099 &  3.79 & 0.32 &   0.16 &    0.23 &        ... &         ... \\
  $\alpha$ CenA &  5792 &  4.31 &  0.26 &  0.13 &     0.19 &          0.17 &           0.20 \\
  $\alpha$ CenB &  5231 &  4.53 &  0.22 &  0.20 &     0.25 &          0.30 &           0.32 \\
   $\tau$ Cet &  5414 &  4.49 & -0.49 &   -0.47 &    -0.42 &         -0.42 &          -0.38 \\
   $\delta$ Eri &  4954 &  3.76 & 0.06 &  -0.02 &     0.02 &          0.00 &           0.03 \\
   $\epsilon$ Eri &  5076 &  4.61 & -0.09 &  -0.11 &    -0.08 &         -0.08 &          -0.06 \\
   $\epsilon$ For &  5123 &  3.45 & -0.60 &  -0.61 &    -0.57 &         -0.65 &          -0.60 \\
   $\beta$ Gem &  4858 &  2.90 & 0.13 &  -0.04 &     0.03 &        ... &         ... \\
   $\xi$ Hya &  5044 &  2.87 & 0.16 &  -0.04 &     0.06 &        ... &         ... \\
   $\beta$ Hyi &  5873 &  3.98 & -0.04 &  -0.24 &    -0.18 &         -0.22 &          -0.19 \\
  Procyon &  6554 &  4.00 & 0.01 &  -0.24 &    -0.19 &        ... &         ... \\
   $\alpha$ Tau &  3927 &  1.11 & -0.37 &  -0.20 &    -0.36 &        ... &         ... \\
   $\beta$ Vir &  6083 &  4.10 & 0.24 &   0.00 &     0.05 &         -0.02 &           0.03 \\
   $\epsilon$ Vir &  4983 &  2.77 & 0.15 &  -0.05 &     0.05 &        ... &         ... \\
  HD22879 &  5868 &  4.27 & -0.86 &  -0.90 &   -0.85 &         -0.95 &          -0.86 \\
  HD49933 &  6635 &  4.20 & -0.41 &  -0.54 &    -0.48 &        ... &         ... \\
  HD84937 &  6356 &  4.06 & -2.03 &  -2.37 &    -2.32 &         -2.32 &          -2.25 \\
 HD102200 &  6155 &  4.22 & -1.12 &  -1.31 &    -1.27 &         -1.34 &          -1.26 \\
 HD106038 &  6121 &  4.55 & -1.25 &  -1.46 &    -1.41 &        ... &         ... \\
 HD107328 &  4496 &  2.09 & -0.33 &  -0.39 &    -0.33 &        ... &         ... \\
 HD122563 &  4635 &  1.40 & -2.75 &  -2.84 &    -2.53 &        ... &         ... \\
 HD140283 &  5792 &  3.65 & -2.29 &  -2.58 &    -2.50 &         -2.59 &          -2.44 \\
 HD201891 &  5948 &  4.30 & -0.97 &  -1.10 &    -1.06 &         -1.15 &          -1.09 \\
 HD298986 &  6223 &  4.19 & -1.26 &  -1.48 &    -1.42 &         -1.49 &          -1.41 \\
      Sun &  5771 &  4.44 & 0.03 &  -0.04 &     0.02 &         -0.04 &          0.00 \\
\hline
\end{tabular}
\end{minipage}
\end{table*}

\begin{table}
\caption{Abundance corrections for \ion{Fe}{i} lines (relative to 1D LTE).}
\label{table:Fe-diff}
\centering
\begin{tabular}{c c c c}
\hline\hline
Star name & 1D NLTE & $\td$ LTE & $\td$ NLTE \\
\hline
    18Sco &     0.04 &          0.03 &           0.08 \\
 Arcturus &     0.04 &        ... &         ... \\
    $\mu$ Ara &     0.07 &          0.04 &           0.06 \\
   $\eta$ Boo &     0.07 &        ... &         ... \\
  $\alpha$ CenA &     0.06 &          0.04 &           0.07 \\
  $\alpha$ CenB &     0.05 &          0.10 &           0.12 \\
   $\tau$ Cet &     0.05 &          0.05 &           0.09 \\
   $\delta$ Eri &     0.04 &          0.02 &           0.05 \\
   $\epsilon$ Eri &     0.03 &          0.03 &           0.05 \\
   $\epsilon$ For &     0.04 &         -0.04 &           0.01 \\
   $\beta$ Gem &     0.07 &        ... &         ... \\
   $\xi$ Hya &     0.10 &        ... &         ... \\
   $\beta$ Hyi &     0.06 &          0.02 &           0.05 \\
  Procyon &     0.05 &        ... &         ... \\
   $\alpha$ Tau &     -0.16 &        ... &         ... \\
   $\beta$ Vir &     0.05 &         -0.02 &           0.03 \\
   $\epsilon$ Vir &     0.10 &        ... &         ... \\
  HD22879 &     0.05 &         -0.05 &           0.04 \\
  HD49933 &     0.06 &        ... &         ... \\
  HD84937 &     0.05 &          0.05 &           0.12 \\
 HD102200 &     0.04 &         -0.03 &           0.05 \\
 HD106038 &     0.05 &        ... &         ... \\
 HD107328 &     0.06 &        ... &         ... \\
 HD122563 &     0.31 &        ... &         ... \\
 HD140283 &     0.08 &         -0.01 &           0.14 \\
 HD201891 &     0.04 &         -0.05 &           0.01 \\
 HD298986 &     0.06 &         -0.01 &           0.07 \\
      Sun &     0.06 &          0.00 &           0.04 \\
\hline
Average & 0.05 & 0.01 & 0.06 \\
\hline
\end{tabular}
\end{table}

\subsection{Ca abundances} \label{sec:ca_abund}

\begin{figure*}[h!]
    \centering
    \includegraphics[width=\linewidth]{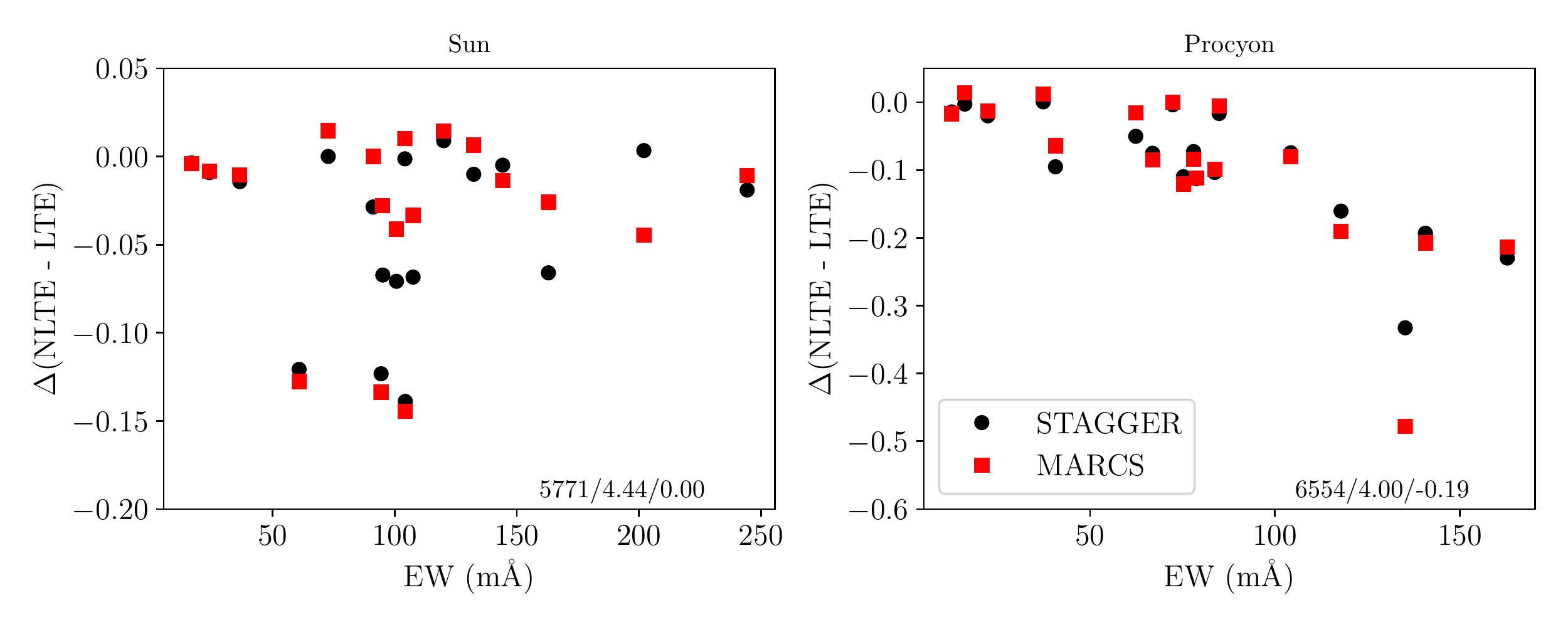}
    \caption{Each panel shows the difference between the line abundances derived from a NLTE and an LTE fit for both $\td$ (Stagger) and MARCS model atmospheres for each \ion{Ca}{i} line fit for the Sun and Procyon as a function of equivalent width of the line. Fits using MARCS model atmospheres are shown as red squares and those using $\td$ model atmospheres are shown as black circles.}
    \label{fig:diff-ca-vs-EW}
\end{figure*}

In addition to Fe, we derived Ca abundances using the new TS to compare our results with what has been observed in the literature. Ca was chosen, because, in contrast to Fe, the lines of \ion{Ca}{i} show pronounced negative NLTE effects. We studied abundances for each individual line given in Table~\ref{table:Ca_lines} of the same stars for which we conducted an individual line analysis for Fe by minimizing the $\chi^2$ value for each line individually in our fitting procedure. Figure~\ref{fig:diff-ca-vs-EW} shows the difference between the NLTE and LTE abundances as a function of EW of a given line for the Sun and Procyon for both MARCS and average Stagger model atmospheres. We chose to present the abundances as a function of EW instead of excitation potential as done in the previous section, because these lines all had excitation potentials between 1.8 and 3~eV, with most lines having an excitation potential $\sim$2.5~eV. In general, we find that NLTE abundances are lower than LTE for both 1D and $\td$ model atmospheres. This result is similar to what was observed by \citet{Mashonkina2017} for the same lines as the ones we used. We also find a correlation with EW for the LTE abundance fits as the abundances of lines with EW $>$ 75 m\AA\ increase with increasing EW. We do not find the same correlation in the NLTE fits. The same correlation for LTE fits was also observed by \citet{Mashonkina2017} in their measurements of Procyon. This behavior explains the trend observed in the $\Delta$(NLTE $-$ LTE) versus EW diagram for Procyon (Fig.~\ref{fig:diff-ca-vs-EW}, right panel), where the difference becomes more negative with increasing EW especially for EW $\gtrsim$ 100~m\AA.

\begin{figure}[h!]
    \centering
    \includegraphics[width=\linewidth]{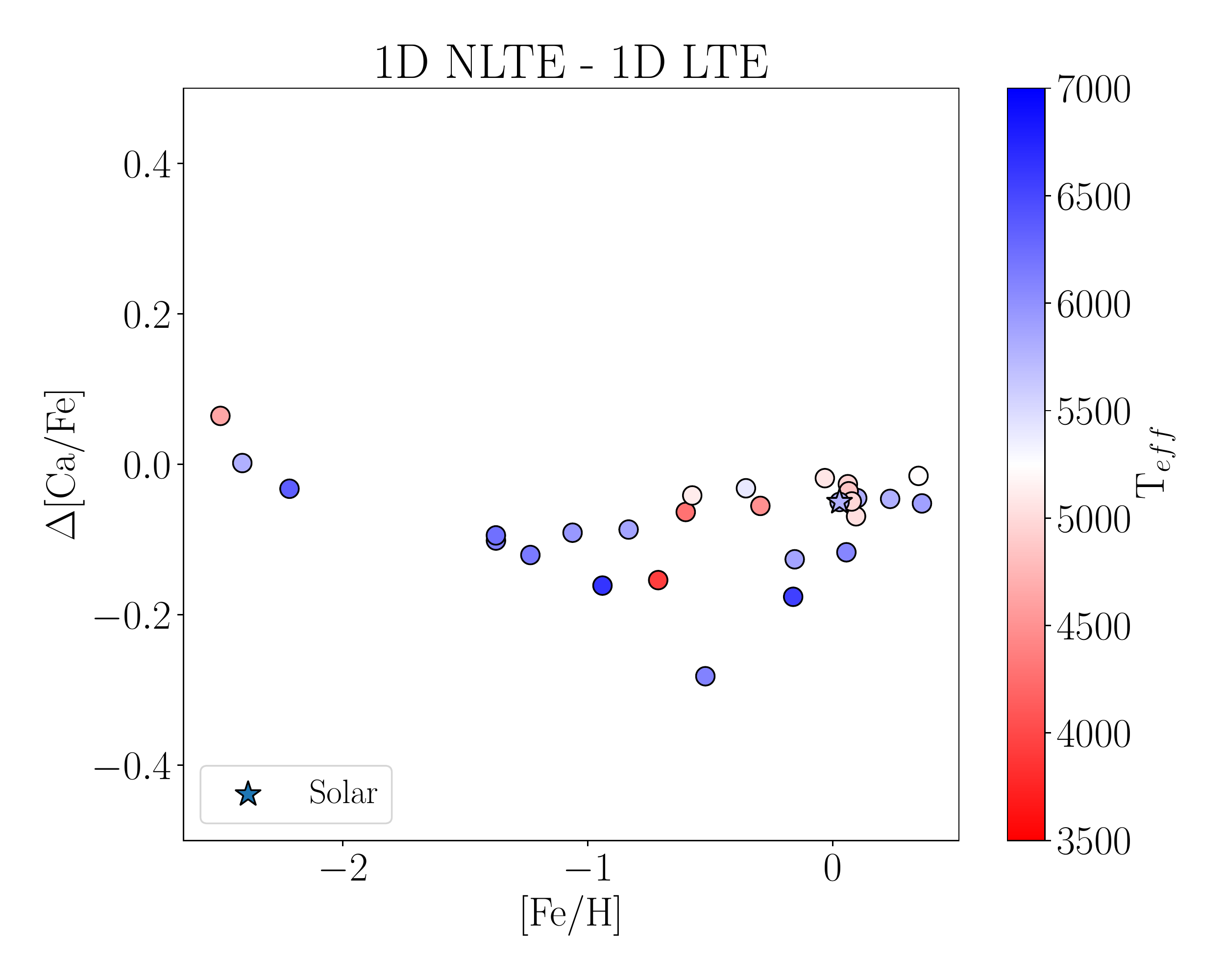}
    \includegraphics[width=\linewidth]{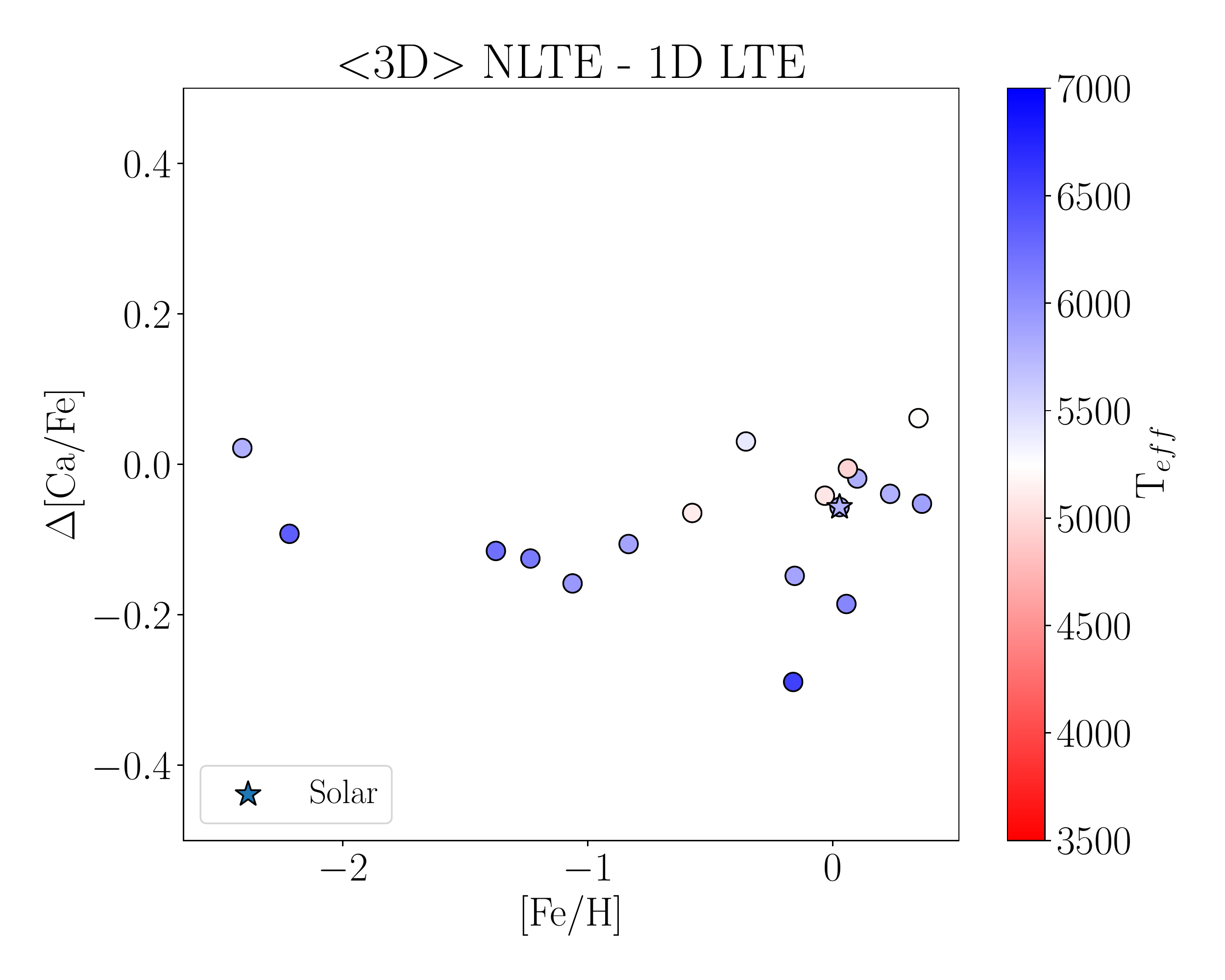}
    \caption{Difference in the [Ca/Fe] value determined using 1D NLTE (top) and $\td$ NLTE (bottom) and 1D LTE as a function of [Fe/H]. Symbols are color coded based on the effective temperature of the star. Fits to spectra of the Sun are indicated as stars.}
    \label{fig:1dnlte-lte-ca}
\end{figure}

For the remaining stars in the sample, we calculated Ca abundances by minimizing the $\chi^2$ for all Ca lines at once similar to what was done for Fe. We used the effective temperature and surface gravity from the literature (and given in Table~\ref{table:Fe}) and our final [Fe/H] value from the previous section. For simplicity, we use the $\td$ NLTE [Fe/H] value for all Ca fits where available, and the 1D NLTE [Fe/H] value for the cases where we were not able to interpolate a $\td$ model as described in the previous section. Our Ca abundance results for the sample of Gaia benchmark stars are shown in Table \ref{table:Ca}. We also analyzed the difference between NLTE and LTE fits to compare with what has been modeled and determined in the literature. We find that the NLTE Ca abundance for both 1D and $\td$ is $\sim$0.1~dex less than the LTE 1D fit with differences ranging from 0.0~dex to roughly $-0.2$~dex as shown in Figure \ref{fig:1dnlte-lte-ca}. For stars with low metallicity ([Fe/H] $\sim-2.0$), the difference is very small and close to zero for most stars in the sample. The smaller difference in these stars is likely due to the mentioned LTE abundance trend with EW. Since the low metallicity stars do not have lines stronger than 100~m\AA, the LTE Ca abundances are not skewed towards stronger values by these lines, which means the results are closer to the NLTE abundances.
Our differences between the 1D LTE, $\td$ LTE, 1D NLTE, and $\td$ NLTE Ca abundance determinations are shown fully in Table \ref{table:Ca}.


\begin{table}
\caption{\ion{Ca}{i} lines used for fitting.}
\label{table:Ca_lines}
\centering
\begin{tabular}{c c r c r}
\noalign{\smallskip}\hline\noalign{\smallskip}
$\lambda$ & $E_{\rm low}$ & $\log gf$ & vdW\tablefootmark{$a$} 
& Ref.\tablefootmark{$b$}
\\
(\AA) & (eV) & & & \\
\noalign{\smallskip}\hline\noalign{\smallskip}
5260.387 & 2.521 & $-$1.719 & 421.260  & 1 \\ 
5261.704 & 2.521 & $-$0.579 & 421.260  & 1 \\ 
5512.980 & 2.933 & $-$0.464 & $-$7.316 & 2 \\ 
5581.965 & 2.523 & $-$0.555 & 400.282  & 1 \\ 
5588.749 & 2.526 &    0.358 & 400.282  & 1 \\ 
5590.114 & 2.521 & $-$0.571 & 399.282  & 1 \\ 
5867.562 & 2.933 & $-$1.570 & $-$7.460 & 2 \\ 
6102.723 & 1.879 & $-$0.850 & 876.233  & 3 \\ 
6122.217 & 1.886 & $-$0.380 & 876.234  & 3 \\ 
6156.023 & 2.521 & $-$2.521 & 978.257  & 4 \\ 
6162.173 & 1.899 & $-$0.170 & 876.234  & 3 \\ 
6166.439 & 2.521 & $-$1.142 & 976.257  & 1 \\ 
6169.042 & 2.523 & $-$0.797 & 976.257  & 1 \\ 
6169.563 & 2.526 & $-$0.478 & 978.257  & 1 \\ 
6439.075 & 2.526 & $ $0.390 & 366.242  & 1 \\ 
6455.598 & 2.523 & $-$1.340 & 365.241  & 2 \\ 
6471.662 & 2.526 & $-$0.686 & 365.241  & 1 \\ 
6493.781 & 2.521 & $-$0.109 & 364.239  & 1 \\ 
6499.650 & 2.523 & $-$0.818 & 364.239  & 1 \\ 
\\
\hline\noalign{\smallskip}
\end{tabular}
\tablefoot{
\tablefoottext{$a$}{Van der Waals broadening parameter from \citet{Anstee1991,Anstee1995} and \citet{Barklem2000}.}
\tablefoottext{$b$}{References for $gf$-values:
(1) \citet{SR} 
(2) \citet{S}
(3) \citet{2009AA...502..989A}
(4) \citet{GESMCHF}
}
}\end{table}

\renewcommand{\footnoterule}{} 
\begin{table*}
\begin{minipage}{\linewidth}
\caption{Results for \ion{Ca}{i} ([Ca/Fe] using a solar abundance of $\log A(\rm Ca) = 6.37$).}
\label{table:Ca}
\centering
\begin{tabular}{c c c c c c c c}
\hline\hline
Star name & Adopted\footnote{See Table \ref{table:Fe} for reference.} $\rightarrow$ & & Derived $\rightarrow$ & \\
& $\teff$ [K] & $\logg$ [cms$^{-2}$] & [Fe/H]\footnote{The $\td$ NLTE fit or 1D NLTE fit from Table \ref{table:Fe} (see Section \ref{sec:ca_abund} for details).} & 1D LTE & 1D NLTE  & $\td$ LTE  & $\td$ NLTE  \\
\hline
    18 Sco &  5810 &  4.44 &    0.07 &   -0.01 &    -0.06 &          0.03 &          -0.03 \\
 Arcturus &  4286 &  1.60 &   -0.63 &    0.11 &     0.04 &       ... &        ... \\
    $\mu$ Ara &  5902 &  4.30 &    0.33 &    0.00 &    -0.05 &          0.03 &          -0.05 \\
   $\eta$ Boo &  6099 &  3.79 &   0.23 &    0.08 &    -0.20 &       ... &        ... \\
  $\alpha$ CenA &  5792 &  4.31 &    0.20 &   -0.01 &    -0.06 &          0.03 &          -0.05 \\
  $\alpha$ CenB &  5231 &  4.53 &    0.32 &   -0.06 &    -0.08 &          0.01 &          -0.00 \\
   $\tau$ Cet &  5414 &  4.49 &   -0.38 &    0.12 &     0.09 &          0.19 &           0.15 \\
   $\delta$ Eri &  4954 &  3.76 &    0.03 &   -0.03 &    -0.05 &         -0.00 &          -0.03 \\
   $\epsilon$ Eri &  5076 &  4.61 &   -0.06 &    0.05 &     0.04 &          0.03 &           0.01 \\
   $\epsilon$ For &  5123 &  3.45 &   -0.60 &    0.35 &     0.31 &          0.39 &           0.28 \\
   $\beta$ Gem &  4858 &  2.90 &    0.03 &    0.01 &    -0.02 &       ... &        ... \\
   $\xi$ Hya &  5044 &  2.87 &    0.06 &    0.01 &    -0.06 &       ... &        ... \\
   $\beta$ Hyi &  5873 &  3.98 &   -0.19 &    0.07 &    -0.05 &          0.08 &          -0.08 \\
  Procyon &  6554 &  4.00 &   -0.19 &    0.07 &    -0.11 &         -0.04 &          -0.22 \\
   $\alpha$ Tau &  3927 &  1.11 &   -0.36 &    0.22 &     0.07 &       ... &        ... \\
   $\beta$ Vir &  6083 &  4.10 &    0.03 &    0.09 &    -0.03 &          0.07 &          -0.09 \\
   $\epsilon$ Vir &  4983 &  2.77 &    0.05 &   -0.03 &    -0.08 &       ... &        ... \\
  HD22879 &  5868 &  4.27 &   -0.86 &    0.27 &     0.19 &          0.23 &           0.17 \\
  HD49933 &  6635 &  4.20 &   -0.48 &    0.13 &    -0.03 &       ... &        ... \\
  HD84937 &  6356 &  4.06 &   -2.25 &    0.33 &     0.30 &          0.34 &           0.24 \\
 HD102200 &  6155 &  4.22 &   -1.26 &    0.28 &     0.16 &          0.23 &           0.15 \\
 HD106038 &  6121 &  4.55 &   -1.41 &    0.28 &     0.18 &       ... &        ... \\
 HD107328 &  4496 &  2.09 &   -0.33 &    0.00 &    -0.05 &       ... &        ... \\
 HD122563 &  4635 &  1.40 &   -2.53 &   -0.62 &    -0.55 &       ... &        ... \\
 HD140283 &  5792 &  3.65 &   -2.44 &    0.09 &     0.09 &          0.05 &           0.11 \\
 HD201891 &  5948 &  4.30 &   -1.09 &    0.25 &     0.15 &          0.21 &           0.09 \\
 HD298986 &  6223 &  4.19 &   -1.41 &    0.24 &     0.15 &          0.21 &           0.13 \\
      Sun &  5771 &  4.44 &    0.00 &    0.02 &    -0.03 &          0.03 &          -0.03 \\
\hline
\end{tabular}
\end{minipage}
\end{table*}

\begin{table}
\caption{Abundance corrections for \ion{Ca}{i} lines (relative to 1D LTE).}
\label{table:Ca-diff}
\centering
\begin{tabular}{c c c c}
\hline\hline
Star name & 1D NLTE & $\td$ LTE & $\td$ NLTE \\
   & dex & dex & dex \\
\hline
    18 Sco &   -0.05 &          0.04 &          -0.02 \\
 Arcturus &    -0.07 &       ... &        ... \\
    $\mu$ Ara &    -0.05 &          0.03 &          -0.05 \\
   $\eta$ Boo &   -0.28 &       ... &        ... \\
  $\alpha$ CenA &      -0.05 &          0.04 &          -0.04 \\
  $\alpha$ CenB &      -0.02 &          0.07 &           0.06 \\
   $\tau$ Cet &   -0.03 &          0.07 &           0.03 \\
   $\delta$ Eri &      -0.02 &          0.03 &           0.00 \\
   $\epsilon$ Eri &    -0.01 &         -0.02 &          -0.04 \\
   $\epsilon$ For &    -0.04 &          0.04 &          -0.07 \\
   $\beta$ Gem &    -0.03 &       ... &        ... \\
   $\xi$ Hya &    -0.07 &       ... &        ... \\
   $\beta$ Hyi &    -0.12 &          0.01 &          -0.15 \\
  Procyon &    -0.18 &         -0.11 &          -0.29 \\
   $\alpha$ Tau &    -0.15 &       ... &        ... \\
   $\beta$ Vir &    -0.12 &         -0.02 &          -0.18 \\
   $\epsilon$ Vir &    -0.05 &       ... &         ... \\
  HD22879 &    -0.08 &         -0.04 &          -0.10 \\
  HD49933 &    -0.16 &       ... &        ... \\
  HD84937 &    -0.03 &          0.01 &          -0.09 \\
 HD102200 &    -0.12 &         -0.05 &          -0.13 \\
 HD106038 &    -0.10 &       ... &        ... \\
 HD107328 &    -0.05 &       ... &        ... \\
 HD122563 &     0.07 &        ... &         ... \\
 HD140283 &     0.00 &         -0.04 &           0.02 \\
 HD201891 &    -0.10 &         -0.04 &          -0.16 \\
 HD298986 &    -0.09 &         -0.03 &          -0.11 \\
      Sun &    -0.05 &          0.01 &          -0.05 \\
\hline
Average & -0.07 & 0.0 & -0.08 \\
\hline
\end{tabular}
\end{table}

\section{Discussion and conclusions} \label{sec:discussion}


We have presented a new version of Turbospectrum capable of handling computations of synthetic spectra using NLTE physics. This new capability plus the existing capability for Turbospectrum to handle $\td$ model atmospheres means that Turbospectrum can now be used to compute model spectra in 1D LTE, 1D NLTE, $\td$ LTE, and $\td$ NLTE. We compared the calculation of equivalent widths of Ca line profiles in NLTE using Turbospectrum to a similar calculation using MULTI (a code well established in the literature for its ability to handle NLTE calculations), and found that the equivalent widths using NLTE profiles agreed within 1 percent for all test cases. This result indicates that the computations by Turbospectrum using NLTE physics are correct, or at least, match well with those from a previous verified source in the literature that has been used to successfully model various observed stellar spectral features in NLTE.

As part of this update, the new version of Turbospectrum is able to compute H line profiles in NLTE as well as line profiles of multiple elements in NLTE at the same time.
Our tests with H line computations also show the importance of $\td$ NLTE calculations when evaluating the temperature of a star using the wings of H lines in the Balmer series. We also provide an example of a spectrum computed using departure coefficient grids for H, O, Mg, Ca, and Fe at the same time to show how multiple elements can be computed in NLTE at once. This capability will be extremely important in future studies that want to model blends of lines from two elements in NLTE, or to model NLTE lines in a crowded spectral region. 

Finally, we provide an example of how the new version of Turbospectrum can be used to determine elemental abundances in observations of stellar spectra by fitting Fe and Ca lines in a number of Gaia FGK benchmark stars. As part of this example, we provide code that involves a Python wrapper that can be used to automate generating spectra with the Fortran based Turbospectrum code, and automates the fitting procedure of spectral features. Our code also includes a script that cross matches information in an LTE Turbospectrum line list with a model atom to create an NLTE formatted Turbospectrum line list. Our abundance values from the example fits agree well with literature values, and show some of the 1D NLTE and $\td$ NLTE effects on abundances determined through spectral fitting as a function of metallicity.

\begin{acknowledgements}
EM and MB acknowledge support by the Collaborative Research centre SFB 881 (projects A5, A10), Heidelberg University, of the Deutsche Forschungsgemeinschaft (DFG, German Research Foundation). 
BP is  supported in part by the Centre National d'Etudes Spatial es (CNES) in the frame of a PLATO grant. This project has received funding from the European Research Council (ERC) under the European Union’s Horizon 2020 research and innovation programme (Grant agreement No. 949173). TO and UH acknowledge support from the Swedish National Space Agency (SNSA/Rymdstyrelsen).
\end{acknowledgements}

%
%


\newpage
\bibliographystyle{aa}
\bibliography{lit,GES_refs}

\end{document}